\documentclass[11pt]{article}

\usepackage[T1]{fontenc}
\usepackage[centertags]{amsmath}
\usepackage[mediummath]{nccmath}
\usepackage{amsfonts}
\usepackage{amssymb}
\usepackage[mathcal]{euscript}
\usepackage{tikz-feynman} 
\tikzfeynmanset{compat=1.0.0} 
\usetikzlibrary{shapes.geometric,calc}
\usepackage[pdftex]{hyperref}
\usepackage{graphicx}
\usepackage[numbers,sort&compress]{natbib}

\usepackage{paperinitial}

\paperinitialization{15mm}{15mm}{15mm}{15mm}{2pt}{10pt}

\DeclareMathOperator{\re}{Re}
\DeclareMathOperator{\im}{Im}

\DeclareMathOperator{\Sp}{Sp}

\newcommand{\lan}{\langle}
\newcommand{\ran}{\rangle}
\newcommand{\cev}[1]{\reflectbox{\ensuremath{\vec{\reflectbox{\ensuremath{#1}}}}}}
\newcommand{\bs}{\boldsymbol}

\newcommand{\e}{\varepsilon}
\newcommand{\vf}{\varphi}
\newcommand{\vk}{\varkappa}
\newcommand{\s}{\sigma}

\newcommand{\al}{\alpha}
\newcommand{\be}{\beta}
\newcommand{\ga}{\gamma}
\newcommand{\Ga}{\Gamma}
\newcommand{\de}{\delta}

\newcommand{\la}{\lambda}

\newcommand{\spx}{\mathbf{x}}

\newcommand{\spp}{\mathbf{p}}
\newcommand{\spq}{\mathbf{q}}
\newcommand{\spk}{\mathbf{k}}

\newcommand{\vcdots}{\rotatebox[origin=c]{-90}{$\cdots$}}

\begin{document}
\allowdisplaybreaks[4]
\frenchspacing
\setlength{\unitlength}{1pt}

\title{{\Large\textbf{Plasmon-polaritons on a single electron}}}

\date{}

\author{
I.M. Akimov${}^{1)}$\thanks{E-mail: \texttt{ima8908@mail.ru}},\;
P.O. Kazinski${}^{1)}$\thanks{E-mail: \texttt{kpo@phys.tsu.ru}},\;
and A.A. Sokolov${}^{1),2)}$\thanks{E-mail: \texttt{asokolov@tpu.ru}}\\[0.5em]
{\normalsize ${}^{1)}$ Physics Faculty, Tomsk State University, Tomsk 634050, Russia}\\
{\normalsize ${}^{2)}$ Tomsk Polytechnic University, Tomsk 634050, Russia}
}

\maketitle

\begin{abstract}

The explicit expression for the photon polarization operator in the presence of a single electron is found in the $in$-$in$ formalism in the one-loop approximation out of the photon mass-shell. This polarization operator describes the dielectric permittivity of a single electron wave packet in coherent scattering processes. The plasmons and plasmon-polaritons supported by a single electron wave packet are described. The two limiting cases are considered: the wavelength of the external electromagnetic field is much smaller than the typical scale of variations of the electron wave packet and the wavelength of the external electromagnetic field is much larger than the size of the electron wave packet. In the former case, there are eight independent plasmon-polariton modes. In the latter case, the plasmons boil down to the dynamical dipole moment attached to a point electron. Thus, in the infrared limit, the electron possesses a dynamical electric dipole moment manifesting itself in coherent scattering processes.

\end{abstract}

\section{Introduction}

Recently, it has been shown in \cite{pra103,KazSol2022,KazSol2023,KRS2023,radet} that there are coherent processes in quantum electrodynamics (QED) where the electron wave packet participates as some kind of a charged fluid. The properties of this fluid are determined by the properties of the electron wave packet. In particular, it was shown in \cite{KazSol2022,KazSol2023} that, in the coherent Compton scattering, the electron wave packet can be endowed with the permittivity tensor so that the photon is scattered by an effective medium with this dielectric permittivity even in the case when only a single electron participates in the process. This is in contrast to the incoherent scattering processes where, in a certain approximation, the electron wave packet behaves as a collection of point incoherent scatterers \cite{MarcuseI,PMHK08,Corson2011,CorsPeat11,WCCP16,PanGov18,Remez19,KdGdAR21,Wong21,RoquesCarmes2023,Karnieli2024}. The on-shell expression for the dielectric permittivity of the electron wave packet derived in \cite{KazSol2022,KazSol2023} appears to coincide with the dielectric permittivity of a gas of free electrons. In the present paper, we generalize the results of \cite{KazSol2022,KazSol2023} and obtain the photon polarization operator out of the photon mass-shell in the presence of a single electron. We reveal thereby that a single electron wave packet supports quasiparticles -- the plasmons -- appearing as singularities of the photon polarization operator. These plasmons hybridize with the electromagnetic field resulting in plasmon-polaritons. We show that there are eight independent plasmon-polariton modes on a single electron wave packet and describe their properties.

Usually, the plasmons are defined as quasiparticles describing the collective excitations of charge density in a plasma \cite{TonLang1929,Lindhard1954,Pines1956,Pitarke2007,Bonitzb2015}. In the coherent scattering processes, where the initial and final states of some particles coincide in the interaction picture \cite{pra103,KazSol2022,KazSol2023,KRS2023,radet,BednNaum2021}, the wave packets of these particles participate coherently, i.e., as some kind of fluids. That is why the analogues of plasmons and plasmon-polaritons can be introduced even for the wave packet of a single electron or any other particle interacting with electromagnetic field. In this sense, one may say that the electron is not so elementary but carries additional degrees of freedom revealing in coherent scattering processes.

\begin{figure}[t]
\begin{equation*}
\begin{gathered}
\resizebox{!}{0.8\height}{%
\begin{tikzpicture}[baseline=(current  bounding  box.center)]
\begin{feynman}
        \vertex  (t1);
        \vertex [below =0.5cm of t1] (p1);
        \node [left =0.75cm of p1](au1);
        \node [right =0.75cm of p1](au2);
        \vertex [above =0.4cm of au1](v1){$\vcdots$};
        \vertex [above =0.4cm of au2](v2){$\vcdots$};
		\node [rectangle, draw,below =0.5cm of p1] (d1){$D$};
        \vertex [below =1cm of d1] (f1);
        \vertex [below =0.5cm of f1] (b1);
        \vertex [left =1.75cm of d1] (i1) {$d$};
        \vertex [right =1.75cm of d1] (ii1) {$d$};
        \vertex [left =1.5cm of f1] (i2) {$\psi$};
        \vertex [right =1.5cm of f1] (ii2) {$\psi$};
        \vertex [left =1.5cm of p1] (ip2){$d$};
        \vertex [right =1.5cm of p1] (ipp2){$d$};
		\diagram{
		(i1) -- [photon] (d1);
		(ii1) -- [photon] (d1);
		(i2) -- [fermion] (f1);
		(ii2) -- [fermion] (f1);
		(ip2) -- [photon] (p1);
        (ipp2) -- [photon] (p1);
        (t1) -- [scalar] (p1);
        (p1) -- [scalar] (d1);
        (d1) -- [scalar] (f1);
        (f1) -- [scalar] (b1);
		};
\end{feynman}
\end{tikzpicture}}
\raisebox{-0.9ex}{+}\\
\raisebox{-0.9ex}{+}\left(
\resizebox{!}{0.8\height}{%
\begin{tikzpicture}[baseline=(current  bounding  box.center)]
\begin{feynman}
        \vertex  (t1);
        \vertex [below =0.5cm of t1] (p1);
        \node [left =0.75cm of p1](au1);
        \node [right =0.75cm of p1](au2);
        \vertex [above =0.4cm of au1](v1){$\vcdots$};
        \vertex [above =0.4cm of au2](v2){$\vcdots$};
		\node [rectangle, draw,below =0.5cm of p1] (d1){$D$};
        \vertex [below =1cm of d1] (f1);
        \vertex [below =0.5cm of f1] (b1);
        \vertex [left =1.75cm of d1] (i1) {$d$};
        \vertex [right =1.75cm of d1] (ii1) {$d$};
        \vertex [left =1.5cm of f1] (i2) {$\psi$};
        \vertex [right =1.5cm of f1] (ii2) {$\psi$};
        \vertex [right =0.5cm of f1] (pv2);
        \vertex [right =1cm of f1] (pv1);
        \vertex [left =1.5cm of p1] (ip2){$d$};
        \vertex [right =1.5cm of p1] (ipp2){$d$};
		\diagram{
		(i1) -- [photon] (d1);
		(ii1) -- [photon] (pv1);
        (pv2) -- [photon] (d1);
		(i2) -- [fermion] (f1);
		(ii2) -- [fermion] (f1);
		(ip2) -- [photon] (p1);
        (ipp2) -- [photon] (p1);
        (t1) -- [scalar] (p1);
        (p1) -- [scalar] (d1);
        (d1) -- [scalar] (f1);
        (f1) -- [scalar] (b1);
		};
\end{feynman}
\end{tikzpicture}}
\raisebox{-0.9ex}{+}
\resizebox{!}{0.8\height}{%
\begin{tikzpicture}[baseline=(current  bounding  box.center)]
\begin{feynman}
        \vertex  (t1);
        \vertex [below =0.5cm of t1] (p1);
        \node [left =0.75cm of p1](au1);
        \node [right =0.75cm of p1](au2);
        \vertex [above =0.4cm of au1](v1){$\vcdots$};
        \vertex [above =0.4cm of au2](v2){$\vcdots$};
		\node [rectangle, draw,below =0.5cm of p1] (d1){$D$};
        \vertex [below =1cm of d1] (f1);
        \vertex [below =0.5cm of f1] (b1);
        \vertex [left =1.75cm of d1] (i1) {$d$};
        \node [crossed dot, draw, right =1.75cm of d1] (ii1) {};
        \vertex [left =1.5cm of f1] (i2) {$\psi$};
        \vertex [right =1.5cm of f1] (ii2) {$\psi$};
        \vertex [right =0.5cm of f1] (pv2);
        \vertex [right =1cm of f1] (pv1);
        \vertex [left =1.5cm of p1] (ip2){$d$};
        \vertex [right =1.5cm of p1] (ipp2){$d$};
		\diagram{
		(i1) -- [photon] (d1);
		(ii1) -- [photon] (pv1);
        (pv2) -- [photon] (d1);
		(i2) -- [fermion] (f1);
		(ii2) -- [fermion] (f1);
		(ip2) -- [photon] (p1);
        (ipp2) -- [photon] (p1);
        (t1) -- [scalar] (p1);
        (p1) -- [scalar] (d1);
        (d1) -- [scalar] (f1);
        (f1) -- [scalar] (b1);
		};
\end{feynman}
\end{tikzpicture}}
\raisebox{-0.9ex}{ +}
\resizebox{!}{0.8\height}{%
\begin{tikzpicture}[baseline=(current  bounding  box.center)]
\begin{feynman}
        \vertex  (t1);
        \vertex [below =0.5cm of t1] (p1);
        \node [left =0.75cm of p1](au1);
        \node [right =0.75cm of p1](au2);
        \vertex [above =0.4cm of au1](v1){$\vcdots$};
        \vertex [above =0.4cm of au2](v2){$\vcdots$};
		\node [rectangle, draw,below =0.5cm of p1] (d1){$D$};
        \vertex [below =1.4cm of d1] (f1);
        \vertex [below =0.5cm of f1] (b1);
        \vertex [left =1.75cm of d1] (i1) {$d$};
        \vertex [right =1.75cm of d1] (ii1) {$d$};
        \vertex [left =1.5cm of f1] (i2) {$\psi$};
        \vertex [right =1.5cm of f1] (ii2) {$\psi$};
        \vertex [above =0.7cm of ii2] (ip1) {$d$};
        \vertex [right =0.5cm of f1] (pv2);
        \vertex [right =1cm of f1] (pv1);
        \node [crossed dot, draw, above =0.5cm of pv2] (iip2) {};
        \vertex [left =1.5cm of p1] (ip2){$d$};
        \vertex [right =1.5cm of p1] (ipp2){$d$};
		\diagram{
		(i1) -- [photon] (d1);
		(ii1) -- [photon] (d1);
        (ip1) -- [photon] (pv1);
        (pv2) -- [photon] (iip2);
		(i2) -- [fermion] (f1);
		(ii2) -- [fermion] (f1);
		(ip2) -- [photon] (p1);
        (ipp2) -- [photon] (p1);
        (t1) -- [scalar] (p1);
        (p1) -- [scalar] (d1);
        (d1) -- [scalar] (f1);
        (f1) -- [scalar] (b1);
		};
\end{feynman}
\end{tikzpicture}}
\raisebox{-0.9ex}{+ c.c.}\right)
\end{gathered}
\end{equation*}
\caption{{\footnotesize The diagrams describing the contributions to the inclusive probability to record a photon in stimulated radiation from a single electron in an external electromagnetic field. The leading terms in the coupling constant are only retained and the diagrams obtained from the depicted ones by permutations of the photons lines are not shown for brevity. The vertical dashed line separates the contributions to transition amplitude from their complex conjugate counterparts. In terms of the $in$-$in$ perturbation theory in the Schwinger representation, the right-hand side contains only the plus vertices, whereas the left-hand side comprises only the minus vertices. The square with $D$ denotes the projector defining the measurement of the properties of an escaping photon. The external lines and the lines that end up on the dashed line or in the detector $D$ are on the mass-shell. The crossed dot means the external electromagnetic field. The initial state of the photons is the coherent state with complex amplitude $d_s(\spk)$. The initial state of the electron is specified by the wave function $\psi_s(\spp)$.}}\label{Fig_Diagram}
\end{figure}
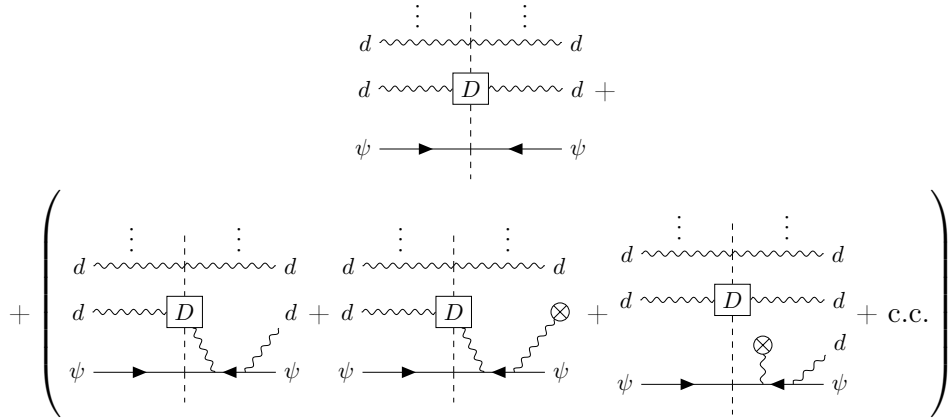

The natural formalism for description of such properties of particle wave packets is the $in$-$in$ perturbation theory \cite{Schw1961,Keld64,CSHY85,GFSh.3,DeWGAQFT.11,Bonitzb2015,CalzHu}. The inclusive probability to detect particles ensuing from the scattering process with a single electron, for example, the radiation from it in the external electromagnetic field, is given by
\begin{equation}\label{P_D_gen}
    P_D=\lan\overline{in}|\hat{S}^\dag \hat{\Pi}_D \hat{S}|\overline{in}\ran,
\end{equation}
in the interaction picture. Here $\hat{\Pi}_D$ is the projector characterizing the detector of escaping particles, $\hat{S}$ is the $S$-matrix, and $|\overline{in}\ran$ is the state of the single electron of a general form. In order to evaluate such a probability, one can employ the $in$-$in$ perturbation theory for the Green's functions. Then applying the reduction formulas to these Green's functions, one can find the probability \eqref{P_D_gen}. The example of such a coherent process is presented in Fig. \ref{Fig_Diagram}. One of the major building blocks of the $in$-$in$ perturbation theory is the exact propagator which is expressed through the polarization (or mass) operator by means of the Schwinger-Dyson equation. In particular, the poles of this propagator specify the quasiparticles of the theory and the contributions of the naive perturbation theory diverge near these poles. From the formal point of view, the aim of the present paper is to describe the poles of the exact photon propagator in the $in$-$in$ perturbation theory with the initial state taken in the form of a single electron wave packet.

It is natural to expect that the properties of the respective quasiparticles resemble the properties of plasmon-polaritons in a gas of noninteracting electrons \cite{Lindhard1954,MelrWeis2003,Silin1960,Tsytovich1961,Melrose2008,VladTysh2011,Melrose2020}. We do establish such a correspondence and, in passing, generalize the results known for a gas of free electrons to the case of arbitrary spin polarized and spatially inhomogeneous states. We also consider the infrared approximation of the photon polarization operator. It turns out that, in this approximation, the photon polarization operator in the presence of the wave packet of a single electron describes a point dynamical electric dipole located at the center of the electron wave packet, the dependence of the photon polarization operator on the form of the electron wave packet disappearing. We derive the action functional that reproduces the effective Maxwell equations in this approximation. Notice that this electric dipole moment is not static, is not aligned along the spin vector as the one appearing in the form-factors of the effective electromagnetic vertex \cite{Bernreuther1991,Hudson2011,PospRitz2014,ACME2018}, and manifests itself in the coherent scattering processes.

The paper is organized as follows. In Sec. \ref{Phot_Pol_Oper}, we introduce the notation and deduce the general formula for the one-loop photon polarization operator in the $in$-$in$ formalism with the initial state taken to be the one electron state of a general form. In Sec. \ref{Effect_Max_Eqs}, we derive the explicit expression for the effective Maxwell equations and consider the approximation for it in the case when the wavelength of the electromagnetic field is much smaller than the typical scale of variations of the electron wave packet. Section \ref{Plasm-Polar_Free_Electr} is devoted to the solutions of the effective Maxwell equations and the description of properties of the plasmon-polaritons. In Sec. \ref{Infrar_Lim}, we consider the infrared limit of the photon polarization operator. In Conclusion, we summarize the results and discuss, as an example, the simple coherent process in QED which is affected by the presence of the plasmon-polaritons on a single electron. In Appendix \ref{App_Green_Func_Bose}, we collect the formulas for the free Green's functions of photons. In Appendix \ref{App_Other_Derivat}, we provide the alternative derivation of the expression for the photon polarization operator. In Appendix \ref{App_G_Z}, we give the complete expressions for the tensors appearing in the one-loop photon polarization operator.

We follow the notation adopted in \cite{KazSol2022,KRS2023}. The Greek indices $\mu$, $\nu$, $\ldots$ are the space-time indices taking the values $\overline{0,3}$ and the Latin $i$, $j$ are the spatial indices. The summation over repeated indices is always understood unless otherwise stated. We also suppose that the quantum states of particles are normalized to unity in some sufficiently large volume $V$. We use the system of units such that $\hbar=c=1$ and $e^2=4\pi\al$, where $\al$ is the fine structure constant. The Minkowski metric $\eta_{\mu\nu}$ is taken with the mostly minus signature.

\section{Photon polarization operator}\label{Phot_Pol_Oper}

The action functional of QED reads
\begin{equation}\label{QED_action}
	\mathcal{L} = -\frac14 F_{\mu\nu}F^{\mu\nu} +\bar{\psi} (i\ga^\mu\partial_\mu-m) \psi -e\bar{\psi} \ga^\mu A_\mu\psi,
\end{equation}
where $m$ is the electron mass, $F_{\mu\nu}:=\partial_{[\mu} A_{\nu]}$ is the electromagnetic field strength tensor, $\psi$ is the Dirac spinor describing the electron-positron quantum field. The action \eqref{QED_action} should also be gauged fixed. Further, we imply the Feynman gauge.

The normalization of mode functions of the electron-positron field and of the electromagnetic field is chosen as in \cite{pra103,KazSol2022,KazSol2023,KRS2023,radet}. In particular,
\begin{equation}
	\hat\psi(x)=\sum_s\int \frac{V d\spp}{(2\pi)^3} \sqrt{\frac{m}{V p_0}} \Bigl[u_s(\spp)e^{-ip_\mu x^\mu}
    \hat{a}_{s}(\spp)+v_s(\spp)e^{i p_\mu x^\mu} \hat{b}_{s}^{\dag}(\spp)\Bigr],
\end{equation}
where the mass-shell condition, $p_0=\sqrt{m^2+\spp^2}$, is fulfilled, $V$ is the normalization volume, and
\begin{equation}\label{spinors}
	u_s(\spp)=\frac{m+\hat p}{\sqrt{2m(p_0+m)}} \genfrac{[}{]}{0pt}{}{\chi_s}{0},\qquad v_s(\spp)
    =\frac{m-\hat p}{\sqrt{2m(p_0+m)}} \genfrac{[}{]}{0pt}{}{0}{\chi_s},
\end{equation}
where the standard representation of $\ga$-matrices is meant. Besides,
\begin{equation}
	(\bs\tau \bs\s) \chi_s=s \chi_s,\qquad s=\pm1.
\end{equation}
The unit vector,
\begin{equation}
	\bs\tau=(\sin\theta\cos\vf,\sin\theta\sin\vf,\cos\theta),
\end{equation}
specifies the direction of the quantization axis of spin projection. There are the relations
\begin{equation}\label{energy_proj}
	\sum_s u_s(\spp)\bar u_s(\spp)=\frac{m+ \hat p}{2m},\qquad\sum_s v_s(\spp)\bar v_s(\spp)=-\frac{m-\hat p}{2m},
\end{equation}
where the hat over the $4$-vector means the convolution of this vector with $\ga_\mu$.

We are interested in finding the polarization operator of a photon in the presence of the wave packet of a single electron. In the interaction picture, the state of the electron at the instant of time $t=0$ takes the form
\begin{equation}\label{one-part_st}
    |\overline{in}\ran=\sqrt{\frac{V}{(2\pi)^3}}\sum_s\int d\spp\vf_s(\spp) \hat{a}^\dag_s(\spp)|0\ran.
\end{equation}
The normalization condition,
\begin{equation}
    \sum_s\int d\spp|\vf_s(\spp)|^2=1,
\end{equation}
is satisfied. For brevity, we consider a pure state of the electron in formula \eqref{one-part_st}. The generalization of formulas to the case of a mixed one-particle electron state is evident and will be given in the final expression for the polarization operator.

In order to obtain the polarization operator, we shall employ the $in$-$in$ formalism \cite{Schw1961,Keld64,CSHY85,GFSh.3,DeWGAQFT.11,Bonitzb2015,CalzHu}. In the Schwinger representation, the interaction action functional becomes
\begin{equation}\label{S_int}
    S_{int}=-e\sum_{a=\pm} a\int d^4x\bar{\psi}^a \ga^\mu A^a_\mu\psi^a.
\end{equation}
The matrix of free propagators of fermions is written as
\begin{equation}
    G^{ab}(x,y)=
    \left[
      \begin{array}{cc}
        S(x,y) & -S_{(-)}(x,y) \\
        S_{(+)}(x,y) & -S_*(x,y) \\
      \end{array}
    \right],
\end{equation}
where
\begin{equation}
\begin{aligned}
    S(x,y)&=-i\lan\overline{in}|T\{\hat{\psi}(x) \hat{\bar{\psi}}(y)\}|\overline{in}\ran,&\qquad
    S_*(x,y)&=i\lan\overline{in}|\tilde{T}\{\hat{\psi}(x) \hat{\bar{\psi}}(y)\}|\overline{in}\ran,\\
    S_{(+)}(x,y)&=-i\lan\overline{in}|\hat{\psi}(x) \hat{\bar{\psi}}(y)|\overline{in}\ran,&\qquad
    S_{(-)}(x,y)&=-i\lan\overline{in}|\hat{\bar{\psi}}(y) \hat{\psi}(x)|\overline{in}\ran.
\end{aligned}
\end{equation}
The operator $\tilde{T}$ means the anti-chronological ordering. These propagators obey the relations
\begin{equation}
    S=S_--S_{(-)}=S_++S_{(+)}=\bar{S}-\frac{i}{2}S_{(1)},\qquad S_*=S_--S_{(+)}=S_++S_{(-)}=\bar{S}+\frac{i}{2}S_{(1)},
\end{equation}
where
\begin{equation}
\begin{aligned}
    S_-(x,y)&=-i\theta(x^0-y^0)\{\hat{\psi}(x),\hat{\bar{\psi}}(y)\}= \theta(x^0-y^0) (S_{(+)}(x,y)+S_{(-)}(x,y)),\\
    S_+(x,y)&=i\theta(y^0-x^0)\{\hat{\psi}(x),\hat{\bar{\psi}}(y)\}= -\theta(y^0-x^0) (S_{(+)}(x,y)+S_{(-)}(x,y)),\\
    \bar{S}(x,y)&=\frac{1}{2}(S_+(x,y)+S_-(x,y)),\qquad S_{(1)}(x,y)=i(S_{(+)}(x,y)-S_{(-)}(x,y)).
\end{aligned}
\end{equation}
Furthermore, there are the symmetry properties,
\begin{equation}
    S_+=\ga^0S_-^\dag \ga^0,\qquad S_*=\ga^0S^\dag \ga^0,\qquad S_{(\pm)}=-\ga^0S^\dag_{(\pm)} \ga^0,
\end{equation}
where the Hermitian conjugation implies not only the transposition of discrete indices but also the transposition of continuous variables. The free propagators for photons are presented in Appendix \ref{App_Green_Func_Bose}.

The functions $S_+$ and $S_-$ are the advanced and retarded Green's functions. They do not depend on the state $|\overline{in}\ran$ and have the standard form \cite{PeskSchr}
\begin{equation}\label{GReen_func_ret}
    S_{\pm}(x,y)=\int\frac{d^4p}{(2\pi)^4}e^{-ip(x-y)}S_{\pm}(p),\qquad S_{\pm}(p)=\frac{\hat{p}+m}{p^2_{\mp}-m^2},
\end{equation}
where $p^\mu_\pm=p^\mu \pm \de^\mu_0i0$. Henceforth, we adopt the agreements for the Fourier transform as in \eqref{GReen_func_ret}. As for the other Green's functions, we deduce
\begin{equation}
\begin{aligned}
    S(x,y)&=S^0(x,y)+i\psi(x)\bar{\psi}(y),&\qquad S_*(x,y)&=S^0_*(x,y)-i\psi(x)\bar{\psi}(y),\\
    S_{(+)}(x,y)&=S^0_{(+)}(x,y)+i\psi(x)\bar{\psi}(y),&\qquad S_{(-)}(x,y)&=S^0_{(-)}(x,y)-i\psi(x)\bar{\psi}(y),
\end{aligned}
\end{equation}
where
\begin{equation}
    \psi(x)=\lan0|\hat{\psi}(x)|\overline{in}\ran=\sum_s\int d\spp\sqrt{\frac{m}{(2\pi)^3 p_0}}u_s(\spp)\vf_s(\spp)e^{-ip_\mu x^\mu}.
\end{equation}
The index $0$ designates the Green's functions defined with respect to the vacuum state. For short, we will refer to such Green's functions and to the contributions coming from them as the vacuum ones. The Fourier transforms of these Green's functions are written as
\begin{equation}
    S^0(p)=\frac{\hat{p}+m}{p^2-m^2+i0},\qquad S^0_*(p)=\frac{\hat{p}+m}{p^2-m^2-i0},\qquad S^0_{(\pm)}(p)=\mp 2\pi i\theta(\pm p_0)\de(p^2-m^2)(\hat{p}+m).
\end{equation}
As for the parts of the Green's functions and the corresponding contributions that depend on the form of the electron wave packet, we will call them the material ones.

The polarization operator is defined in the standard way,
\begin{equation}
    \Pi_{\mu\nu}^{ab}(x,y):=\frac{\de^2\bar{\Ga}[A_a,\bar{\psi}_a,\psi_a]}{\de A^\mu_a(x) \de A^\nu_b(y)} \Big|_{A^\mu_a(x)=0,\psi_a(x)=\bar{\psi}_a(x)=0},
\end{equation}
where $\bar{\Ga}[A_a,\bar{\psi}_a,\psi_a]$ denotes the quantum corrections to the effective action in the $in$-$in$ formalism. The polarization operator is the sum of all the one-particle irreducible diagrams with one incoming and one outgoing photon lines. Keeping in mind the form of the interaction vertex \eqref{S_int}, we obtain in the leading order of perturbation theory (see Appendix \ref{App_Other_Derivat} for the other derivation)
\begin{equation}\label{polar_oper_Schw}
\begin{aligned}
    \Pi^{\mu\nu}_{++}(x,y)&=ie^2\Sp(\ga^\mu S(x,y)\ga^\nu S(y,x)),&\qquad \Pi^{\mu\nu}_{--}(x,y)&=ie^2\Sp(\ga^\mu S_*(x,y)\ga^\nu S_*(y,x)),\\
    \Pi^{\mu\nu}_{+-}(x,y)&=ie^2\Sp(\ga^\mu S_{(-)}(x,y)\ga^\nu S_{(+)}(y,x)),&\qquad \Pi^{\mu\nu}_{-+}(x,y)&=ie^2\Sp(\ga^\mu S_{(+)}(x,y)\ga^\nu S_{(-)}(y,x)).
\end{aligned}
\end{equation}
Notice that
\begin{equation}
    \Pi^{\mu\nu}_{+-}(x,y)=\Pi^{\nu\mu}_{-+}(y,x),\qquad [\Pi^{\mu\nu}_{++}(x,y)]^*=-\Pi^{\mu\nu}_{--}(x,y).
\end{equation}
Splitting the propagators in \eqref{polar_oper_Schw} into the vacuum contribution and the contribution depending on the form of the electron one-particle state \eqref{one-part_st}, we see that the polarization operator can be cast into the form
\begin{equation}\label{Pi0_Pi_psi}
    \Pi^{\mu\nu}_{ab}(x,y)= \overset{0}{\Pi}{}^{\mu\nu}_{ab}(x,y)+ \overset{\psi}{\Pi}{}^{\mu\nu}_{ab}(x,y),
\end{equation}
where $\overset{0}{\Pi}{}^{\mu\nu}_{ab}$ is the vacuum polarization operator. In the one-loop approximation we consider, only $\overset{0}{\Pi}{}^{\mu\nu}_{++}$ and $\overset{0}{\Pi}{}^{\mu\nu}_{--}$ contain ultraviolet divergencies. These divergencies are canceled out by means of the standard renormalization procedure.

Upon renormalization (see, e.g., \cite{PeskSchr}),
\begin{equation}
    \overset{0}{\Pi}{}^{\mu\nu}_{++}(k)=(k^2\eta^{\mu\nu}-k^\mu k^\nu)\overset{0}{\Pi}(k^2+i0),\qquad\overset{0}{\Pi}(k^2)=\frac{2\al}{\pi}\int_0^1 dxx(1-x)\ln\Big(1-x(1-x)\frac{k^2}{m^2}\Big),
\end{equation}
and
\begin{equation}
    \overset{0}{\Pi}{}^{\mu\nu}_{--}(k)=-(k^2\eta^{\mu\nu}-k^\mu k^\nu)\overset{0}{\Pi}(k^2-i0).
\end{equation}
Besides,
\begin{equation}
    \im \overset{0}{\Pi}(k^2\pm i0)=\mp\frac{\al}{3}\sqrt{1-\frac{4m^2}{k^2}} \big(1+\frac{2m^2}{k^2}\big)\theta(k^2-4m^2).
\end{equation}
The rest components of the vacuum polarization tensor turn into
\begin{equation}
\begin{split}
    \overset{0}{\Pi}{}^{\mu\nu}_{+-}(k)=&\frac{2i\al}{3} \sqrt{1-\frac{4m^2}{k^2}} \big(1+\frac{2m^2}{k^2}\big) \theta(-k_0)\theta(k^2-4m^2) (k^2\eta^{\mu\nu}-k^\mu k^\nu),\\
    \overset{0}{\Pi}{}^{\mu\nu}_{-+}(k)=&\frac{2i\al}{3} \sqrt{1-\frac{4m^2}{k^2}} \big(1+\frac{2m^2}{k^2}\big) \theta(k_0)\theta(k^2-4m^2) (k^2\eta^{\mu\nu}-k^\mu k^\nu).
\end{split}
\end{equation}
The renormalized polarization operator \eqref{Pi0_Pi_psi} satisfies the Ward identities
\begin{equation}\label{Ward_ident}
    \partial^x_\mu\Pi^{\mu\nu}_{ab}(x,y)=0,\qquad \partial^y_\nu\Pi^{\mu\nu}_{ab}(x,y)=0.
\end{equation}

For further analysis, it is convenient to switch to the Keldysh representation \cite{Keld64,CSHY85}. The electromagnetic fields in the Schwinger and Keldysh representations are related as
\begin{equation}
    A^\pm_\mu=A_\mu^c \pm\frac12 A^q_\mu.
\end{equation}
Therefore, the polarization operator in the Keldysh representation takes the form
\begin{equation}\label{polar_op_keld}
\begin{split}
    \|\Pi_{pr}\|&=
    \left[
      \begin{array}{cc}
        1 & 1 \\
        \frac12 & -\frac12 \\
      \end{array}
    \right]
    \left[
      \begin{array}{cc}
        \Pi_{++} & \Pi_{+-} \\
        \Pi_{-+} & \Pi_{--} \\
      \end{array}
    \right]
    \left[
      \begin{array}{cc}
        1 & \frac12 \\[0.5em]
        1 & -\frac12 \\
      \end{array}
    \right]=\\
    &=
    \left[
      \begin{array}{cc}
        \Pi_{++} +\Pi_{+-} +\Pi_{-+} +\Pi_{--} & \frac12(\Pi_{++} -\Pi_{+-} +\Pi_{-+} -\Pi_{--}) \\[0.5em]
        \frac12(\Pi_{++} +\Pi_{+-} -\Pi_{-+} -\Pi_{--}) & \frac14(\Pi_{++} -\Pi_{+-} -\Pi_{-+} +\Pi_{--}) \\
      \end{array}
    \right],
\end{split}
\end{equation}
where $p,r\in\{c,q\}$. Substituting the representation \eqref{Pi0_Pi_psi} with renormalized vacuum polarization tensors into this expression, we come to
\begin{equation}\label{Pi_0}
\begin{gathered}
    \overset{0}{\Pi}{}^{\mu\nu}_{qc}(k)=(k^2\eta^{\mu\nu}-k^\mu k^\nu)\overset{0}{\Pi}(k^2_+),\qquad \overset{0}{\Pi}{}^{\mu\nu}_{cq}(k)=(k^2\eta^{\mu\nu}-k^\mu k^\nu)\overset{0}{\Pi}(k^2_-),\\ \overset{0}{\Pi}{}^{\mu\nu}_{qq}(k)=-\frac{4i\al}{3}\sqrt{1-\frac{4m^2}{k^2}}\big(1+\frac{2m^2}{k^2}\big)\theta(k^2-4m^2) (k^2\eta^{\mu\nu}-k^\mu k^\nu),
\end{gathered}
\end{equation}
and
\begin{equation}\label{Pi_psi}
\begin{split}
    \overset{\psi}{\Pi}{}^{\mu\nu}_{qc}(x,y)&=-e^2\bar{\psi}(x)\ga^\mu S_-(x,y) \ga^\nu\psi(y) -e^2\bar{\psi}(y)\ga^\nu S_+(y,x) \ga^\mu \psi(x),\\
    \overset{\psi}{\Pi}{}^{\mu\nu}_{cq}(x,y)&=-e^2\bar{\psi}(x)\ga^\mu S_+(x,y) \ga^\nu\psi(y) -e^2\bar{\psi}(y)\ga^\nu S_-(y,x) \ga^\mu \psi(x),\\
    -i\overset{\psi}{\Pi}{}^{\mu\nu}_{qq}(x,y)&=-e^2\bar{\psi}(x)\ga^\mu\psi(x) \bar{\psi}(y)\ga^\nu\psi(y) +\frac{e^2}{2}\bar{\psi}(x)\ga^\mu \overset{(0)}{S}_{(1)}(x,y) \ga^\nu\psi(y) +\frac{e^2}{2}\bar{\psi}(y)\ga^\nu \overset{(0)}{S}_{(1)}(y,x) \ga^\mu \psi(x),
\end{split}
\end{equation}
and also
\begin{equation}
    \Pi^{\mu\nu}_{cc}(x,y)=0.
\end{equation}
The fact that $\Pi^{\mu\nu}_{cc}=0$ is a general property of the effective action in the $in$-$in$ formalism in the Keldysh representation. It is a consequence of causality \cite{CSHY85}.

Generally, causality implies
\begin{equation}
    \frac{\de^n\Ga[\phi_c,\phi_q]}{\de\phi^{A_1}_c(x_1)\cdots \de\phi^{A_n}_c(x_n)}\Big|_{\phi^A_q=0}=0,\quad\text{i.e.,}\quad \Ga[\phi_c,\phi_q]\Big|_{\phi^A_q=0}=0,
\end{equation}
where $\phi^A_p(x)$ denotes all the fields of the theory. The quadratic part of the effective action defining the properties of quasiparticles in the theory looks as
\begin{equation}\label{quadr_part}
    \frac{\de^2\Ga[\phi_c,\phi_q]}{\de\phi^{A_1}_p(x_1)\de\phi^{A_2}_r(x_2)}\Big|_{\phi^A_q=0}=
    \left[
      \begin{array}{cc}
        0 & \Ga_{cq} \\
        \Ga_{qc} & \Ga_{qq} \\
      \end{array}
    \right],
\end{equation}
where $\Ga_{pr}$ denotes the corresponding second derivatives of the effective action. The inverse to the matrix operator \eqref{quadr_part} is the exact propagator in the $in$-$in$ perturbation theory in the Keldysh representation. The null vectors of \eqref{quadr_part} characterize the states of quasiparticles. It is clear that the column $(w_c,0)$, where
\begin{equation}\label{reterded_eff_eqs}
    \Ga_{qc} w_c=0,
\end{equation}
gives such solutions. Therefore, it order to describe the quasiparticles, it is sufficient to find the solutions to the effective equations \eqref{reterded_eff_eqs}. We will find them in Sec. \ref{Plasm-Polar_Free_Electr}. Evidently, the poles of the exact propagator in the Keldysh and Schwinger representations are the same.

In the case of a mixed one-particle electron state, expressions \eqref{Pi_psi} are modified in an obvious manner. One just needs to replace
\begin{equation}
    \psi(x)\bar{\psi}(y)\rightarrow\rho(x,y),
\end{equation}
where the relativistic density matrix has been introduced,
\begin{equation}
    \rho(x,y):=m\sum_{s,s'}\int \frac{d\spp d\spp'}{(2\pi)^3} \frac{u_s(\spp)\bar{u}_{s'}(\spp')}{\sqrt{p_0p'_0}} \rho_{ss'}(\spp,\spp')e^{-ip_\mu x^\mu +ip'_\mu y^\mu},
\end{equation}
and $\rho_{ss'}(\spp,\spp')$ is the density matrix in the momentum representation obeying the normalization condition
\begin{equation}
    \sum_s\int d\spp \rho_{ss}(\spp,\spp)=1.
\end{equation}
For example,
\begin{equation}
    -i\overset{\psi}{\Pi}{}^{\mu\nu}_{qq}(x,y)=-e^2\Sp[\ga^\mu\rho(x,y)\ga^\nu\rho(y,x)] +\frac{e^2}{2}\Sp[\ga^\mu \overset{(0)}{S}_{(1)}(x,y) \ga^\nu\rho(y,x)] +\frac{e^2}{2}\Sp[\ga^\nu \overset{(0)}{S}_{(1)}(y,x) \ga^\mu \rho(x,y)].
\end{equation}
In the same way, the other matrix elements of polarization operator \eqref{Pi_psi} are written in terms of the density matrix.

\section{Effective Maxwell equations}\label{Effect_Max_Eqs}

Let us derive the explicit expression for the effective Maxwell equations describing the electromagnetic fields in the presence of the single electron wave packet. As we shall see, the expression for the effective Maxwell equations is greatly simplified in the case of an electron wave packet that is narrow in momentum space. This will allow us to find the solutions of the effective Maxwell equations explicitly.

The effective Maxwell equations in the momentum representation can be cast into the form
\begin{equation}\label{Max_eqn_eff}
    -(1-\overset{0}{\Pi}(k'^2_+))(k'^2\eta^{\mu\nu} -k'^\mu k'^\nu)A_\nu(k') +\int \frac{d^4k}{(2\pi)^4} \overset{\psi}{\Pi}{}^{\mu\nu}_{qc}(k',k) A_\nu(k)=0,
\end{equation}
where the transition into the momentum representation is defined as
\begin{equation}
    \Pi_{qc}(k',k)=\int dxdy e^{ik'x-iky}\Pi_{qc}(x,y).
\end{equation}
The material part of the polarization operator entering into the effective Maxwell equations is convenient to write as
\begin{equation}\label{Pi_psi_ex1}
    \overset{\psi}{\Pi}{}^{\mu\nu}_{qc}(k',k)=-2\pi e^2m\sum_{s,s'} \int d\spp_c d\spq\de(k'-k-q) \frac{\rho_{ss'}(\spp,\spp')}{\sqrt{p_0p'_0}} \bar{u}_{s'}(\spp')\Big[\frac{\ga^\mu(\hat{p}_c+\hat{k}_c+m)\ga^\nu}{(p^c+k^c_+)^2-m^2} +\frac{\ga^\nu(\hat{p}_c-\hat{k}_c+m)\ga^\mu}{(p^c-k^c_+)^2-m^2} \Big]u_s(\spp),
\end{equation}
where $q_\mu=k'_\mu-k_\mu=p_\mu-p'_\mu$, $p_\mu^c:=(p_\mu+p'_\mu)/2$, and $k_\mu^c:=(k_\mu+k'_\mu)/2$. Notice that the following relations hold
\begin{equation}
\begin{gathered}
    (p_cq)=0,\qquad (pp')=m^2-q^2/2,\qquad p_c^2=m^2-q^2/4,\\
    (p_c-k_c)^2-m^2=k_c^2 -2(p_ck_c)-q^2/4,\qquad (p_c+k_c)^2-m^2=k_c^2 +2(p_ck_c)-q^2/4.
\end{gathered}
\end{equation}
One can verify by explicit calculations that the polarization operator \eqref{Pi_psi_ex1} obeys the Ward identities \eqref{Ward_ident}, viz., it is orthogonal to the vectors $k_\nu$ and $k'_\mu$. In order to single out the dependence of \eqref{Pi_psi_ex1} on the electron spin, we use the representation (see, for details, \cite{KRS2023})
\begin{equation}\label{GZ_decomp}
    \bar{u}_{s'}(\spp')\Big[\frac{\ga^\mu(\hat{p}_c+\hat{k}_c+m)\ga^\nu}{(p^c+k^c_+)^2-m^2} +\frac{\ga^\nu(\hat{p}_c-\hat{k}_c+m)\ga^\mu}{(p^c-k^c_+)^2-m^2} \Big]u_s(\spp)=\de_{s's}G^{\mu\nu}(\spp,\spp') -(\s_a)_{s's} \tau_{ai} Z^{i\mu\nu}(\spp,\spp'),
\end{equation}
where the set of vectors, $\tau^i_a$, has been introduced. These vectors can be regarded as a tetrad that maps the vector on the Poincar\'{e} sphere into the spin vector in the $x$-space,
\begin{equation}
\begin{gathered}
    \bs{\tau}_1=\re\mathbf{f},\qquad \bs{\tau}_2=\im\mathbf{f},\qquad\bs{\tau}_3=\bs\tau,\\
    \mathbf{f}=e^{i\vf}(\cos\theta \cos\vf-i\sin\vf,\cos\theta \sin\vf+i\cos\vf,-\sin\theta).
\end{gathered}
\end{equation}
The coefficients in the expansion \eqref{GZ_decomp} are expressed in terms of the traces
\begin{equation}
\begin{split}
    G^{\mu\nu}(\spp,\spp') =&\frac{1}{4m\sqrt{(p_0+m)(p'_0+m)}}\times\\ &\times\Sp\Big[\frac{\ga^\mu(\hat{p}_c+\hat{k}_c+m)\ga^\nu(\hat{p}+m)\frac{1+\ga^0}{2}(\hat{p}'+m)}{(p^c+k^c_+)^2-m^2} +\frac{\ga^\nu(\hat{p}_c-\hat{k}_c+m)\ga^\mu(\hat{p}+m)\frac{1+\ga^0}{2}(\hat{p}'+m)}{(p^c-k^c_+)^2-m^2} \Big],\\
    Z^{i\mu\nu}(\spp,\spp') =&\frac{-1}{4m\sqrt{(p_0+m)(p'_0+m)}}\times\\
    &\times\Sp\Big[\frac{\ga^\mu(\hat{p}_c+\hat{k}_c+m)\ga^\nu(\hat{p}+m) \ga^i\ga^5 \frac{1+\ga^0}{2}(\hat{p}'+m)}{(p^c+k^c_+)^2-m^2} +\frac{\ga^\nu(\hat{p}_c-\hat{k}_c+m)\ga^\mu(\hat{p}+m) \ga^i\ga^5 \frac{1+\ga^0}{2}(\hat{p}'+m)}{(p^c-k^c_+)^2-m^2} \Big].
\end{split}
\end{equation}
These tensors are orthogonal to $k_\nu$ and $k'_\mu$ as the polarization operator is. The traces of $\gamma$-matrices appearing in the above formulas are readily evaluated. However, the resulting expressions are rather huge and are removed to Appendix \ref{App_G_Z}.

It is useful to introduce the Weyl symbol of the polarization operator. The following relations hold
\begin{equation}
\begin{gathered}
    \Pi(k',k)=\int d^4x_c e^{i(k'-k)x_c} \Pi(x_c,k_c),\qquad \Pi(x,y)=\int \frac{d^4k_c}{(2\pi)^4} e^{-ik_c(x-y)} \Pi(x_c,k_c),\\ \Pi(x_c,k_c)=\int\frac{d^4q}{(2\pi)^4} e^{-iqx_c}\Pi(k_c+q/2,k_c-q/2)=\int d^4z e^{ik_c z}\Pi(x_c+z/2,x_c-z/2),
\end{gathered}
\end{equation}
where $\Pi(x_c,k_c)$ is the Weyl symbol and $x_c:=(x+y)/2$. Then the action of the material part of the polarization operator on the vector potential is written as
\begin{equation}\label{Pi_psi_A}
    \int d^4y\overset{\psi}{\Pi}{}^{\mu\nu}_{qc}(x,y) A_\nu(y)=\Big[ \exp(\frac{i}{2}\frac{\vec{\partial}}{\partial x^\rho} \frac{\vec{\partial}}{\partial k_\rho} ) \overset{\psi}{\Pi}{}^{\mu\nu}_{qc}(x,k) \Big]  \exp(i \frac{\cev{\partial}}{\partial k_\s} \frac{\vec{\partial}}{\partial x^\s} ) A_\nu(x)\Big|_{k=0},
\end{equation}
in the coordinate representation. Let $l$ be the typical scale of variations of the Weyl symbol of the polarization operator $\overset{\psi}{\Pi}_{qc}(x,k)$ with respect to the variable $x$, whereas $\vk$ is its typical scale of variations with respect to the variable $k$. If
\begin{equation}\label{smooth_variation}
    \vk l\gg1,
\end{equation}
then
\begin{equation}
    \exp(\frac{i}{2}\frac{\vec{\partial}}{\partial x^\rho} \frac{\vec{\partial}}{\partial k_\rho} ) \overset{\psi}{\Pi}{}^{\mu\nu}_{qc}(x,k)\approx \overset{\psi}{\Pi}{}^{\mu\nu}_{qc}(x,k),
\end{equation}
and the contribution \eqref{Pi_psi_A} becomes
\begin{equation}
    \int d^4y\overset{\psi}{\Pi}{}^{\mu\nu}_{qc}(x,y) A_\nu(y)\approx\overset{\psi}{\Pi}{}^{\mu\nu}_{qc}\big(x,i\frac{\partial}{\partial z}\big) A_\nu(x+z)\Big|_{z=0}.
\end{equation}
In this approximation, the effective Maxwell equations \eqref{Max_eqn_eff} in the coordinate representation are brought into the form
\begin{equation}
    (\Box\eta^{\mu\nu} -\partial^\mu\partial^\nu)A_\nu(x)+ \overset{\psi}{\Pi}{}^{\mu\nu}_{qc}\big(x,i\frac{\partial}{\partial z}\big) [1-\overset{0}{\Pi}(-\Box^z_+)]^{-1} A_\nu(x+z)\Big|_{z=0}=0.
\end{equation}
The contribution of the vacuum polarization is convenient to take into account by introducing the effective charge
\begin{equation}\label{effective_charge}
    e^2_{eff}(k^2):=e^2/(1-\overset{0}{\Pi}(k^2_+)).
\end{equation}
Then the effective Maxwell equations become
\begin{equation}\label{Max_eqn_eff_app}
    (\Box\eta^{\mu\nu} -\partial^\mu\partial^\nu)A_\nu(x)+ \overset{\psi}{\Pi}{}^{\mu\nu}_{qc}\big(x,i\frac{\partial}{\partial z}\big)\Big|_{e^2\rightarrow e^2_{eff}} A_\nu(x+z)\Big|_{z=0}=0.
\end{equation}
Below we shall obtain the solutions to these equations and provide the estimates for the parameters $l$ and $\vk$.

The Weyl symbol of the polarization operator \eqref{Pi_psi_ex1} is given by
\begin{equation}\label{Pi_psi_Weyl}
    \overset{\psi}{\Pi}{}^{\mu\nu}_{qc}(x,k_c)=-e^2 m\sum_{s,s'} \int \frac{d\spp_c d\spq}{(2\pi)^3} \frac{\rho_{ss'}(t;\spp_c+\spq/2,\spp_c-\spq/2)}{\sqrt{p_0p'_0}} e^{i\spq\spx} \big[\de_{s's}G^{\mu\nu}(\spp,\spp') -(\s_a)_{s's} \tau_{ai} Z^{i\mu\nu}(\spp,\spp') \big],
\end{equation}
where the electron density matrix at the instant of time $t$,
\begin{equation}
    \rho_{ss'}(t;\spp,\spp')=\rho_{ss'}(\spp,\spp')e^{i(p'_0-p_0)t},
\end{equation}
has been introduced. If the electron wave packet, $\rho_{ss'}(\spp,\spp')$, is sufficiently narrow in the momentum space, i.e.,
\begin{equation}
    |\spq|\ll p^c_0,\qquad |\spq|\ll|\spp^c_0|,\qquad |\spq|\ll |k^c_0|,\qquad |\spq|\ll|\spk_c|,
\end{equation}
where $\spp_c^0$ is the typical value of momenta in the wave packet, then one can neglect the dependence of the integrand of \eqref{Pi_psi_Weyl} on $q_\mu$ in the leading order in $|\spq|$ with the exception of dependence of the density matrix on $\spq$. Let us introduce the matrix-valued Wigner function
\begin{equation}
    \rho_{ss'}(x,\spp_c)=\int\frac{d\spq}{(2\pi)^3} e^{i\spq\spx} \rho_{ss'}(t;\spp_c+\spq/2,\spp_c-\spq/2).
\end{equation}
This matrix is Hermitian but not positive-definite, in general. It can be written as
\begin{equation}
    \rho_{ss'}(x,\spp_c)=\frac12 \rho(x,\spp_c) [\de_{ss'} +\xi_a(x,\spp_c) (\s_a)_{ss'}].
\end{equation}
The normalization condition becomes
\begin{equation}
    \int d\spx d\spp_c\rho(x,\spp_c)=1.
\end{equation}
The vector $\xi_a(x,\spp_c)$ characterizes the spin polarization of the wave packet and
\begin{equation}
    \lan\s_a\ran= \sum_{s,s'}\int d\spp \rho_{ss'}(\spp,\spp) (\s_a)_{s's}=\sum_{s,s'}\int d\spx d\spp_c \rho_{ss'}(x,\spp_c) (\s_a)_{s's}=\int d\spx d\spp_c\rho(x,\spp_c)\xi_a(x,\spp_c).
\end{equation}
If the wave packet is not polarized, then $\xi_a(x,\spp_c)=0$. As long as the Wigner function is not positive-definite, $\rho(x,\spp_c)$ is not nonnegative, in general, and the modulus of the real vector $\xi_a(x,\spp_c)$ can be larger than unity.

With the aid of the Wigner function, the Weyl symbol of the polarization operator \eqref{Pi_psi_Weyl} can be written as
\begin{equation}\label{Pi_psi_Weyl_sml_rec}
    \overset{\psi}{\Pi}{}^{\mu\nu}_{qc}(x,k_c)=-e^2m \int \frac{d\spp_c}{p_c^0} \rho(x,\spp_c) \big[G^{\mu\nu}(\spp_c,\spp_c) -\xi_{i}(x,\spp_c) Z^{i\mu\nu}(\spp_c,\spp_c) \big],
\end{equation}
where
\begin{equation}
    \xi_{i}(x,\spp_c)= \xi_{a}(x,\spp_c)\tau_{ai}.
\end{equation}
In the small recoil limit, $|\spq|\rightarrow0$, it is not difficult to obtain that
\begin{equation}\label{G_Z_appr}
\begin{split}
    G^{\mu\nu}(\spp_c,\spp_c)=&-\frac{(k_cp_c)^2}{m((k_cp_c)^2 -k_c^4/4)} \Big[\eta^{\mu\nu} -\frac{k_c^{(\mu} p_c^{\nu)}}{(k_cp_c)} +\frac{k_c^2 p_c^\mu p_c^\nu}{(k_cp_c)^2} \Big],\\
    Z^{i\mu\nu}(\spp_c,\spp_c)=& \frac12\frac{i k_c^2}{(k_cp_c)^2-k_c^4/4} \Big[\e^{i\mu\nu\rho} k^c_\rho +\frac{p^i_c}{m}\e^{\mu\nu\rho\s}k^c_\rho \frac{p^c_\s+m\de^0_\s}{p^c_0+m} \Big].
\end{split}
\end{equation}
It is clear that $Z^{i\mu\nu}$ is antisymmetric with respect to $\mu$ and $\nu$, while $G^{\mu\nu}$ is symmetric. These tensors satisfy the Ward identities, i.e., they are orthogonal to $k_\mu^c$ with respect to the indices $\mu$ and $\nu$. In the small recoil limit, on the photon mass-shell, $k_c^2=0$, the tensor $Z^{i\mu\nu}$ vanishes, whereas the expression for $G^{\mu\nu}$ coincides with the one derived in the papers \cite{KazSol2022,KazSol2023}. Notice that
\begin{equation}
    \xi_i(x,\spp_c)Z^{i\mu\nu}(\spp_c,\spp_c)=\frac12\frac{i k_c^2}{(k_cp_c)^2-k_c^4/4}s_\s(x,\spp_c)\e^{\s\mu\nu\rho} k^c_\rho,
\end{equation}
where
\begin{equation}\label{smu_zeta}
    s^\mu=\Big(\frac{(\bs{\xi}\spp_c)}{m},\bs{\xi}+\frac{\spp_c(\bs{\xi}\spp_c)}{m(p^c_0+m)}\Big),\qquad s^\mu p^c_\mu=0,\qquad s^2=-\bs{\xi}^2.
\end{equation}
As a result, the polarization operator \eqref{Pi_psi_Weyl_sml_rec} is brought into an explicitly Lorentz covariant form. Substituting  \eqref{G_Z_appr} into \eqref{Pi_psi_Weyl_sml_rec}, we see that the typical scale of variations of the Weyl symbol of the polarization operator with respect to the variable $k_c$ is of order $m$, i.e., $\vk=m$. The scale $l$ is of order of the typical scale of variations of the Wigner function of the wave packet, $\rho_{ss'}(x,\spp_c)$, with respect to the variable $x$. Consequently, the condition \eqref{smooth_variation} is fulfilled when $ml\gg1$.

Suppose additionally that the wave packet possesses a definite polarization at any point in the $x$-space, i.e., in virtue of narrowness of $\rho(x,\spp_c)$ in the momentum space, $\xi_i(x,\spp_c)$ can be taken at the point $\spp_c=\spp_c^0$. Introducing the Lorentz-covariant probability density,
\begin{equation}
    \rho(x):=m\int \frac{d\spp_c}{p^0_c} \rho(x,\spp_c),
\end{equation}
we write the polarization operator as
\begin{equation}\label{polar_op_app}
    \overset{\psi}{\Pi}{}^{\mu\nu}_{qc}(x,k_c)=\frac{\omega_p^2(x)}{(k_cp_c)^2-k_c^4/4} \big[(k_cp_c)^2\eta^{\mu\nu} -(k_cp_c)k_c^{(\mu} p_c^{\nu)} +k_c^2p_c^\mu p_c^\nu-\frac{imk_c^2}{2}\e^{\mu\nu\rho\s}k^c_\rho s_\s(x,\spp_c) \big],
\end{equation}
where $\omega_p^2(x):=e^2\rho(x)/m$ and $\spp_c=\spp_c^0$. As it has been already mentioned, the vacuum polarization is taken into account in the effective Maxwell equations by substituting the effective charge $e^2_{eff}(k_c^2)$ instead of $e^2$. We will denote the corresponding plasma frequency as $\omega_p(x,k_c)$. It is clear from the above derivation that if the density matrix of the wave packet $\rho_{ss'}(\spp,\spp')$ is strongly localized near several distinct values of momenta $(\spp_c^0)_k$, where $k$ enumerates these values, then the polarization operator is approximately the sum of expressions of the form \eqref{polar_op_app} corresponding to the contributions from those values of momenta. A similar expression for the polarization operator arises in the quantum multistream model of a cold plasma (see, e.g., \cite{Melrose2008,Haas2000}).

\section{Plasmon-polaritons on a single free electron}\label{Plasm-Polar_Free_Electr}

In this section, we shall obtain the solutions of the effective Maxwell equations for the electromagnetic field in the presence of the electron wave packet. It turns out that the wave packet of a single electron supports the quasiparticles -- plasmons. This leads, in turn, to the existence of plasmon-polaritons -- the solutions of the Maxwell equation in the presence of a single electron wave packet. We shall describe the dynamics of these quasiparticles.

The polarization operator \eqref{polar_op_app} and the effective Maxwell equations \eqref{Max_eqn_eff_app} are explicitly Lorentz covariant. Therefore, it is convenient to analyze these equations in the electron rest frame, where $\spp_c^0=0$ and
\begin{equation}
    p_c^\mu=(m,0),
\end{equation}
and then to make a Lorentz boost to the reference frame required. In what follows in this section, for brevity, we do not write the index $0$ at $\spp_c^0$. Moreover, we will not write the index $c$ at $p^c_\mu$ and $k^c_\mu$. The polarization operator \eqref{polar_op_app} has singularities when
\begin{equation}\label{disp_law_plasm}
    (kp)^2-k^4/4=m^2k_0^2-k^4/4=0.
\end{equation}
As is known (see, e.g., \cite{WeinbergB.12}), the singularities in the effective action appear when there are new degrees of freedom in the theory -- quasiparticles. In the case at issue, it is natural to call these quasiparticles plasmons. The condition \eqref{disp_law_plasm} results in the dispersion law of plasmons
\begin{equation}\label{plasm_disp_law}
    k_0=\sqrt{m^2+\spk^2}\pm m.
\end{equation}
The perturbation theory describing coherent scattering of photons by an electron wave packet does not work near these values of momenta $k^\mu$ since the corresponding diagrams diverge. It is necessary to resum them or, what is equivalent, to find the exact solutions to the effective Maxwell equations.

Further in this section, we will consider the case when $\omega_p(x,k)$ and $s^\mu(x,\spp)$ can be regarded as independent of $x$. This situation is realized if the wavelength of a plasmon-polariton is much smaller than the typical scale of variations of $\omega_p(x,k)$ and $s^\mu(x,\spp)$ with respect to the variable $x$. In that case, the effective Maxwell equations become translation invariant and, on performing the Fourier transform, turn into the linear matrix equation
\begin{equation}\label{Max_eqn_eff_const}
    \Big\{-k^2\eta^{\mu\nu} +k^\mu k^\nu +\frac{\omega_p^2(k)}{(kp)^2-k^4/4} \big[(kp)^2\eta^{\mu\nu} -(kp)k^{(\mu} p^{\nu)} +k^2p^\mu p^\nu -\frac{im k^2}{2}\e^{\mu\nu\rho\s}k_\rho s_\s \big]\Big\} A_\nu(k)=0.
\end{equation}
The solutions to this equation are obtained in an obvious way. The dielectric permittivity tensor corresponding to the effective Maxwell equations \eqref{Max_eqn_eff_const} coincides with that was given in formula (3.21) of \cite{Lindhard1954} (see also \cite{MelrWeis2003,Silin1960,Tsytovich1961,Melrose2008,VladTysh2011,Melrose2020}) for a gas of free relativistic electrons. Notice that the contributions due to vacuum polarization and spin of the electron wave packet are not presented in \cite{Lindhard1954}. The spin dependent contribution is given in \cite{MelrWeis2003} in the particular case of helical electron states when $\mathbf{s}\parallel\spp$.

Let us consider, at first, the case of an unpolarized electron wave packet. In that case, $s^\mu=0$ and Eqs. \eqref{Max_eqn_eff_const} are invariant under spatial rotations in the rest frame of the electron. Hence, we may put $k^\mu=(k_0,0,0,k_z)$ without loss of generality. Taking into account the dependence of $\omega_p$ on $k$ perturbatively, we find eight branches of the dispersion law for plasmon-polaritons: the two longitudinal modes
\begin{equation}\label{longitud_modes}
    a)\;A^\mu(k)=(k_z,0,0,k_0),\qquad k_0=\sqrt{2m^2+k_z^2 \pm2m\sqrt{m^2-\omega_p^2+k_z^2}};
\end{equation}
and the six transverse modes
\begin{equation}\label{trans_modes_A}
    A^\mu(k)=(0,\al,\be,0),
\end{equation}
where $\al$ and $\be$ are some complex constants independent of $k_z$. The dispersion law for transverse modes is obtained by solving the equation
\begin{equation}\label{trans_modes}
    b)\;k_0^6 - k_0^4 (4m^2 + 3 k_z^2) + k_0^2 [4m^2 (k_z^2+\omega_p^2) + 3 k_z^4] - k_z^6=0.
\end{equation}
This is a cubic polynomial equation with respect to $k_0^2$. Its solutions can be written explicitly but they are fairly cumbersome. Therefore, we will not present them here. The dependence of $\omega_p$ on $k^2$ is considered perturbatively: $k^2$ is found from the solutions \eqref{longitud_modes} or $\eqref{trans_modes}$ with $\omega_p$ taken without vacuum polarization, i.e., $\omega_p$ is evaluated with the ordinary electron charge, and is substituted into $\omega_p(k^2)$ in the solutions \eqref{longitud_modes} or $\eqref{trans_modes}$. Notice that, as is seen from \eqref{longitud_modes}, the longitudinal modes are the oscillations of the electric field along the $z$ axis in the reference frame we consider.

\begin{figure}[tp]
\centering
\includegraphics*[width=0.49\linewidth]{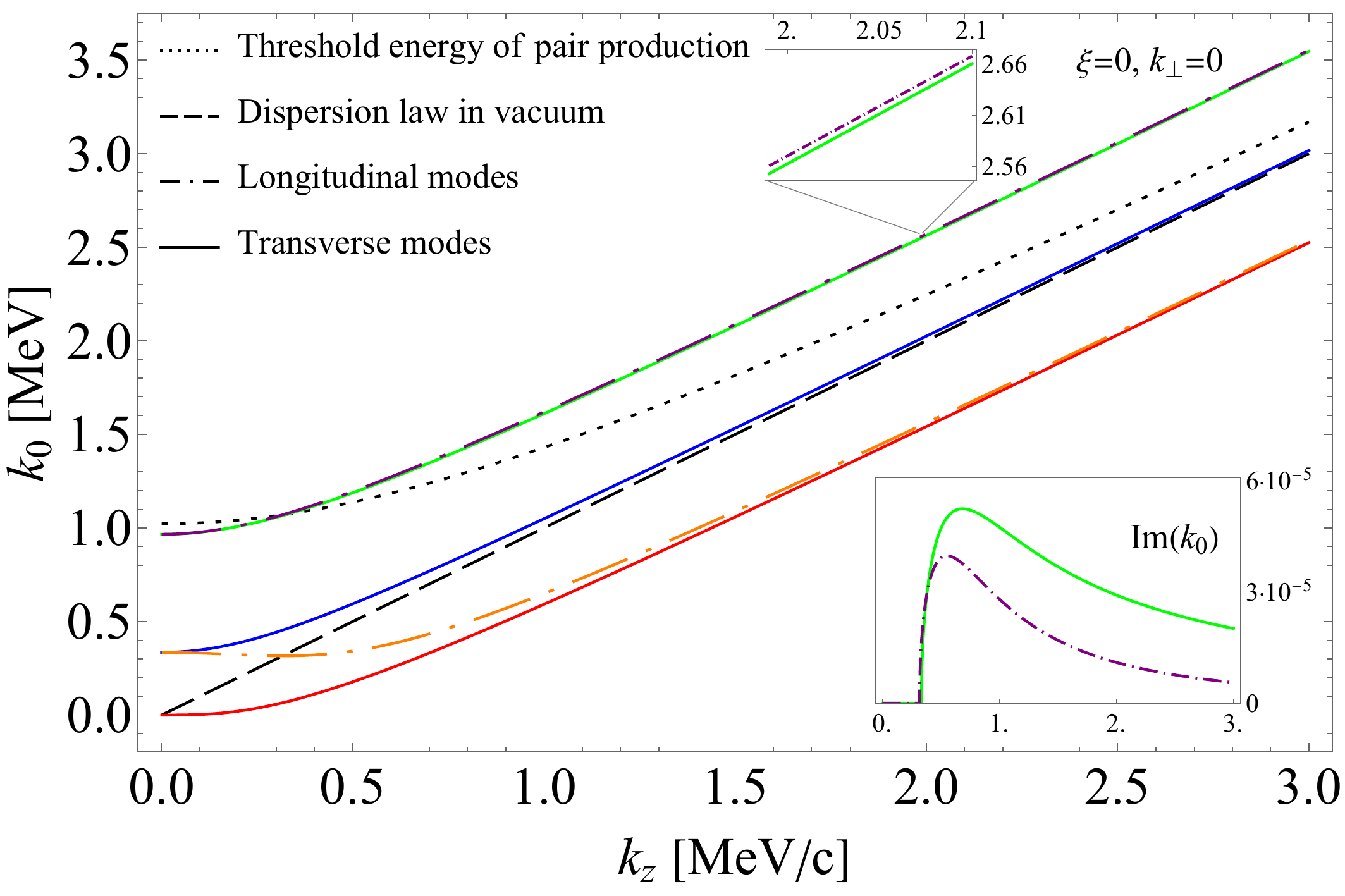}\,
\includegraphics*[width=0.49\linewidth]{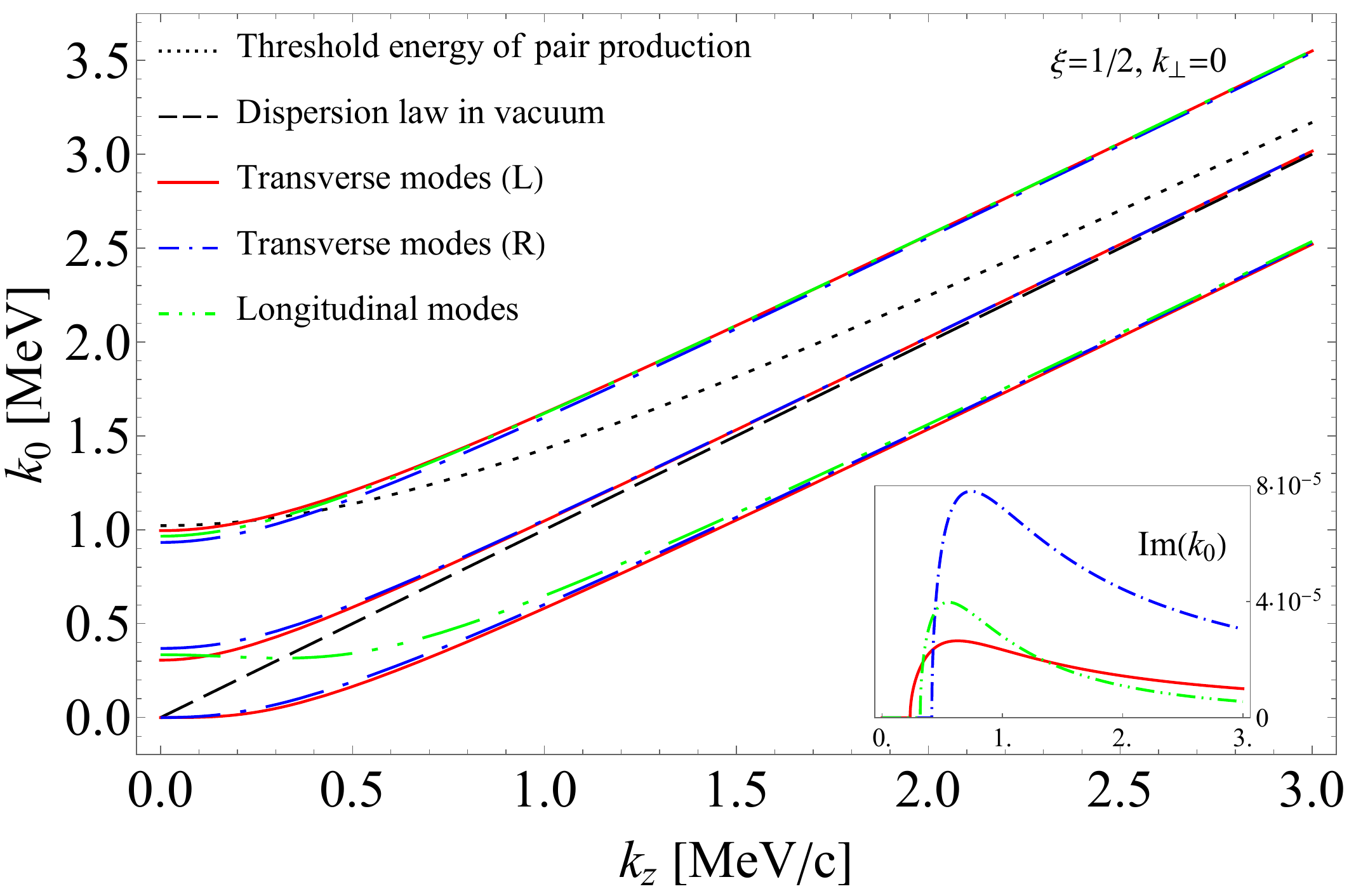}
\caption{{\footnotesize On the left panel: The dispersion laws of plasmon-polariton modes on a single unpolarized electron in rest frame of the electron. The plasma frequency with the ordinary electron charge is taken to be $\omega_p^2/m^2=0.1$ for the curves of the dispersion laws to be discernable. The dispersion law in vacuum and the electro-positron pair creation threshold are also shown. The energies of plasmon-polaritons above the pair creation threshold possess positive imaginary parts presented in the inset. On the right panel: The same as on the left panel but for a single polarized electron with $\xi=1/2$.}}
\label{Fig_Upolarized}
\end{figure}

For $\omega_p\ll m$, the asymptotics of the dispersion laws \eqref{longitud_modes} and \eqref{trans_modes} for small and large momenta are written as
\begin{equation}\label{disp_law_long}
\begin{split}
    a)\; k_0&\approx 2m -\frac{\omega_p^2}{4m} +\frac{k_z^2}{2m} \big(1+\frac{3\omega_p^2}{8m^2}\big)+O(k_z^3)\approx k_z+m+O(1/k_z),\\
    k_0&\approx \omega_p\big(1 -\frac{k_z^2}{4m^2}\big)+O(k_z^3)\approx k_z-m+O(1/k_z),
\end{split}
\end{equation}
and
\begin{equation}\label{disp_law_trans}
\begin{gathered}
    b)\; k_0\approx 2m -\frac{\omega_p^2}{4m} +\frac{k_z^2}{2m}\big(1+\frac{\omega_p^2}{4m^2}\big)+O(k_z^3)\approx k_z+m+O(1/k_z), \\
    k_0\approx\omega_p +\frac{k_z^2}{2\omega_p}+O(k_z^3)\approx k_z+O(1/k_z),\qquad k_0\approx\frac{k_z^3}{2m\omega_p}+O(k_z^4)\approx k_z-m+O(1/k_z).
\end{gathered}
\end{equation}
Notice that for the electron wave packet whose spatial dimensions are of order of the Bohr radius, $1/(\al m)$, the plasma frequency
\begin{equation}
    \omega_p/m\sim 4\pi\al^4\approx 3.6\times 10^{-8}.
\end{equation}
The plots of the dispersion laws \eqref{longitud_modes} and \eqref{trans_modes} are presented in Fig. \ref{Fig_Upolarized}. The modes with dispersion laws given on the first lines of formulas \eqref{disp_law_long}, \eqref{disp_law_trans} become unstable due to possible creation of the electron-positron pairs when
\begin{equation}
    k^2\geqslant 4m^2\;\Rightarrow\;k_z^2\gtrsim \omega_p^2.
\end{equation}
In this case, $\omega_p^2$ entering into the dispersion laws contains an imaginary part coming from the imaginary part of the effective charge $e^2_{eff}(k^2)$. Note, however, that this imaginary part is quite small in comparison with the real part of the energy of plasmon-polaritons.

Now let us consider the wave packet of a polarized electron. As above, we will solve the effective Maxwell equations \eqref{Max_eqn_eff_const} in the rest frame of the electron so that
\begin{equation}
    s^\mu=(0,0,0,\xi),\qquad |\xi|\leqslant1,
\end{equation}
where $\xi$ is a constant quantity characterizing a degree of spin polarization of the electron wave packet. In order to simplify the resulting expressions, we firstly consider the plasmon-polaritons propagating along the $z$ axis. Then, as in the case of an unpolarized electron, there are the longitudinal and transverse modes: the two longitudinal modes, which are left intact and have the form \eqref{longitud_modes}; and the six transverse modes,
\begin{equation}\label{trans_modes_polar}
    A^\mu(k)=(0,1,\mp i,0),
\end{equation}
with the dispersion law determined by the equation
\begin{equation}\label{trans_modes_polar_disp}
    h_\perp^\mp:=k_0^6 - k_0^4 (4m^2 + 3 k_z^2) \mp 2\xi m\omega^2_p (k_0^3-k_0k_z^2) + k_0^2 [4m^2 (k_z^2+\omega_p^2) + 3 k_z^4] - k_z^6=0.
\end{equation}
The signs in \eqref{trans_modes_polar} and \eqref{trans_modes_polar_disp} are agreed. The dependence of $\omega_p$ on $k^2$ is taken into account as it was described above. It is clear that the modes \eqref{trans_modes_polar} correspond to the plasmon-polaritons with circular polarization. The presence of an electron polarization removes the degeneracy of transverse modes \eqref{trans_modes_A} that exists in the case of an unpolarized electron. For $\xi=0$, Eqs. \eqref{trans_modes_polar_disp} turn into Eq. \eqref{trans_modes} determining the dispersion law of plasmon-polaritons on an unpolarized electron. Equations \eqref{trans_modes_polar_disp} taken with both signs possess the twelve solutions with respect to $k_0$, six of them being nonnegative. The plots of the dispersion laws \eqref{longitud_modes}, \eqref{trans_modes_polar_disp} are presented in Fig. \ref{Fig_Upolarized}. Notice that, in the case $|\xi|\rightarrow1$, the dispersion laws of two transverse modes tend to the dispersion laws of plasmons \eqref{plasm_disp_law}. In particular, they become independent of $\omega_p$ in this limit and so the plasmon-polariton mode corresponding to the dispersion law with the plus sign in \eqref{plasm_disp_law} gets almost stable despite the fact that $k^2\geqslant4m^2$ for this mode.

\begin{figure}[tp]
\centering
\includegraphics*[width=0.49\linewidth]{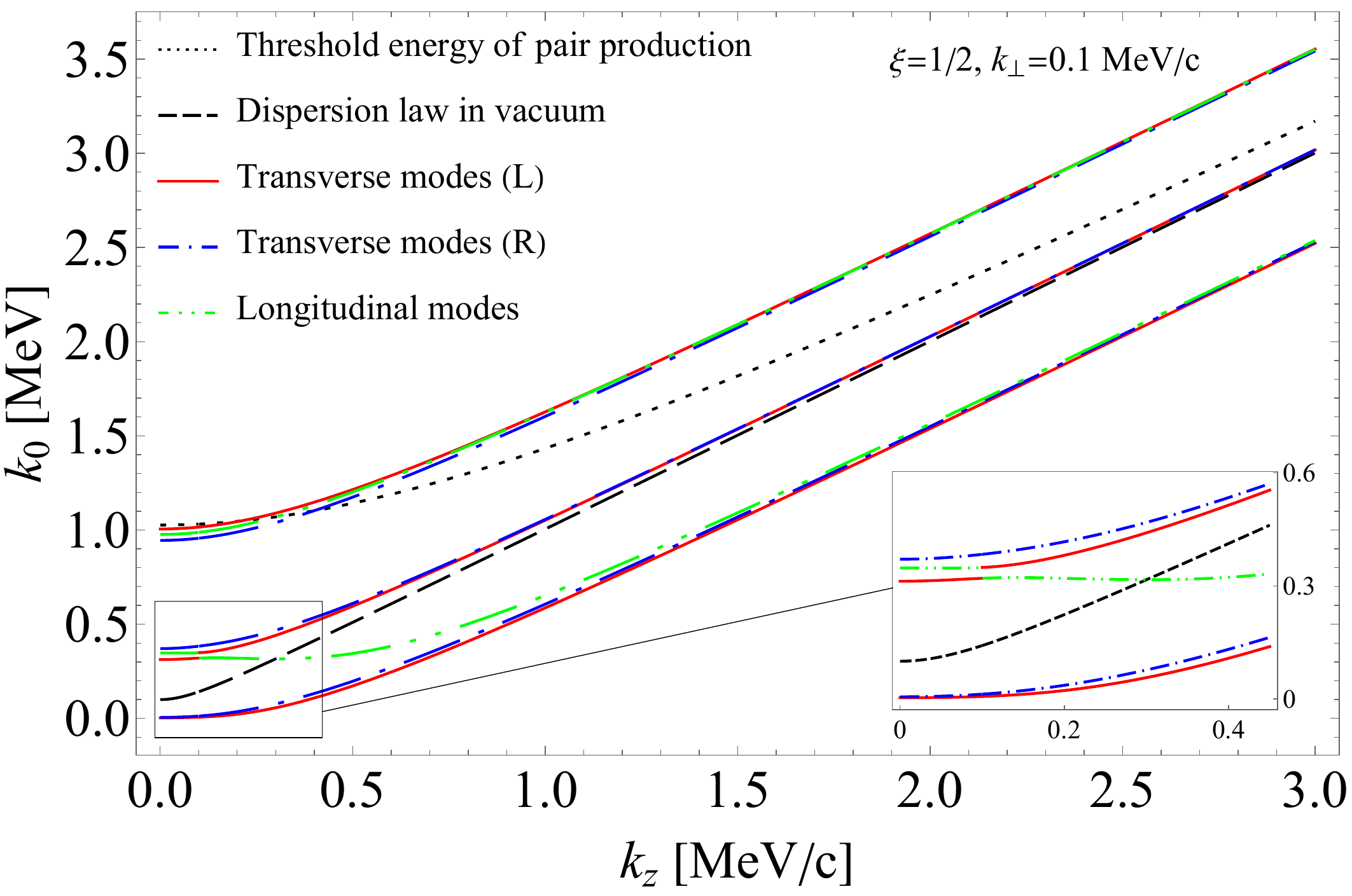}
\caption{{\footnotesize The same as on the right panel in Fig. \ref{Fig_Upolarized} but for $k_\perp=0.1$ MeV/c. It is seen in the inset that the transverse and longitudinal modes intersecting at $k_\perp=0$ separate for $k_\perp\neq 0$ due to their interference. The plasmon-polariton modes do not have a definite polarization (longitudinal or transverse) in the case $k_\perp\neq0$. Their polarizations on the plot above are defined as inherited from the case $k_\perp=0$, i.e., by considering the dispersion law for $k_\perp\neq0$ as a continuous deformation of the dispersion law for $k_\perp\neq0$ with respect to the parameter $k_\perp$.}}
\label{Fig_Gen_Pol}
\end{figure}

In the case when the plasmon-polaritons propagate along an arbitrary direction, the complete set of solutions to \eqref{Max_eqn_eff_const} can also be found. Without loss of generality, we can take
\begin{equation}
    k^\mu=(k_0,0,k_\perp,k_z).
\end{equation}
In that case, the electromagnetic potential is given by
\begin{equation}\label{sol_gen}
    A^\mu(k)=\big( 2\xi \omega_p^2 k_\perp h_\perp,i h_\parallel h_\perp, 2\xi \omega_p^2 k_0(h_\perp-4m^2\omega_p^2 k_z^2), 8\xi k_0k_\perp |k_z| \omega_p^4 \big),
\end{equation}
where
\begin{equation}
\begin{split}
    h_\parallel&:=k_0^4 -2k_0^2(2m^2+\spk^2) +4m^2\omega_p^2 +\spk^4,\\
    h_\perp&:=k_0^6 - k_0^4 (4m^2 + 3 \spk^2) + k_0^2 [4m^2 (\spk^2+\omega_p^2) + 3 \spk^4] - \spk^6,
\end{split}
\end{equation}
and $k_\perp=\sqrt{\spk^2-k_z^2}$. Notice that for $\spk^2=k_z^2$, the equation $h_\parallel=0$ defines the dispersion law of longitudinal modes \eqref{longitud_modes} for an unpolarized electron, whereas the equation $h_\perp=0$ coincides with \eqref{trans_modes}. The dispersion law of the corresponding mode is to be substituted into the solution \eqref{sol_gen}. As in the particular cases considered above, there are eight independent plasmon-polariton modes. Their dispersion laws are specified by the equation
\begin{equation}
    h_\parallel h^+_\perp h^-_\perp +4\xi^2\omega_p^2 k_\perp^2(k_0^2-\spk^2)^2((k_0^2-\spk^2)^2 -4m^2k_0^2)=0.
\end{equation}
This is the polynomial equation of the eight degree with respect to $k_0^2$. As is seen from this equation, if the dispersion laws of different modes intersect at $k_\perp=0$, then for $k_\perp\neq0$ the intersection of these modes disappears because of their interference. The plots of the dispersion laws in this case are given in Fig. \ref{Fig_Gen_Pol}.

\section{Infrared limit}\label{Infrar_Lim}

Consider the other limit of the general expression \eqref{Pi_psi_ex1} for the polarization operator. Assume that the wavelength of the external electromagnetic field is much larger than the typical dimensions of the electron wave packet in the coordinate space. Furthermore, we suppose that the electron wave packet is sufficiently narrow in the momentum space. Formally, these conditions are reduced to
\begin{equation}
    |k^\mu|\ll |p^\mu|, \qquad|q^\mu|\ll |p^\mu|,
\end{equation}
and
\begin{equation}
    \rho_{ss'}(\spp,\spp')\approx \rho_{ss'}(\spp_c,\spp_c) e^{-i\spq \spx_0},
\end{equation}
where $\spx_0$ is the position of the center of the electron wave packet. In this approximation, we have in the leading order
\begin{equation}\label{G_IR}
    G^{\mu\nu}(\spp,\spp')\approx-\frac1{m}\Big[\eta^{\mu\nu} -\frac{k^\mu p^\nu_c}{(kp_c)} -\frac{p^\mu_c k'^\nu}{(k'p_c)} +\frac{(k'k)p^\mu_cp^\nu_c}{(k'p_c)(kp_c)} \Big]=:-\frac1{m}\pi^{\mu\nu}(k',k),
\end{equation}
and $Z^{i\mu\nu}(\spp,\spp')$ turns out to be of higher order of smallness. It is not difficult to check that expression \eqref{G_IR} is orthogonal to $k'_\mu$ and $k_\nu$.

The convolution of the material part of the polarization operator with the electromagnetic potential becomes
\begin{equation}\label{convolution}
    \int \frac{d^4k}{(2\pi)^4} \overset{\psi}{\Pi}{}_{qc}^{\mu\nu}(k',k)A_\nu(k)\approx \frac{e^2}{p^0_c} \int\frac{d\spq}{(2\pi)^3} e^{-i\spq\spx_0} \pi^{\mu\nu}(k',k'-q)A_\nu(k'-q),
\end{equation}
in the approximation we consider. The dependence on the form of the wave packet disappears in this approximation. Bearing in mind that
\begin{equation}
    q^0=p^0-p'^0\approx (\mathbf{v}_c\spq),
\end{equation}
where $\mathbf{v}_c:=\spp_c/p^0_c$, the convolution \eqref{convolution} turns into
\begin{equation}
    \int \frac{d^4k}{(2\pi)^4} \overset{\psi}{\Pi}{}_{qc}^{\mu\nu}(k',k)A_\nu(k)\approx \frac{e^2}{p^0_c} \int dx e^{ik'x} \pi^{\mu\nu}\Big(i\frac{\partial}{\partial x},i\frac{\partial}{\partial x}\Big|_{A}\Big)A_\nu(x)\de(\spx-\spx_0-\mathbf{v}_c x^0),
\end{equation}
where $\partial/\partial x|_{A}$ denotes the derivative acting on the variable $x^\mu$ only in the potential $A_\nu(x)$. Introducing the world line of the center of the wave packet, $x^\mu(\tau)$, where $\tau$ is the natural parameter, we can write
\begin{equation}
    \de(\spx-\spx_0-\mathbf{v}_c x^0)=\int d\tau \frac{p^0_c}{m} \de(x-x(\tau)).
\end{equation}
Then the effective Maxwell equations have the form
\begin{equation}\label{Max_eqn_eff_IR}
    (\Box\eta^{\mu\nu} -\partial^\mu\partial^\nu) A_\nu(x) +\frac{e^2}{m}\int d\tau \pi^{\mu\nu}\Big(i\frac{\partial}{\partial x},i\frac{\partial}{\partial x}\Big|_{A}\Big)\de(x-x(\tau))A_\nu(x)=0.
\end{equation}
Inasmuch as the polarization operator $\pi^{\mu\nu}$ possesses singularities, one needs to define precisely the nonlocal operators $1/(kp_c)$ and $1/(k'p_c)$. As it has been derived in Sec. \ref{Effect_Max_Eqs}, the retarded Green's functions should be taken in the expression for the polarization operator to ensure the causality of the effective Maxwell equations.

One can get rid of the nonlocal contributions to the effective equations \eqref{Max_eqn_eff_IR} by introducing the additional field (the degrees of freedom) such that, on eliminating this field, equations \eqref{Max_eqn_eff_IR} are reproduced for the electromagnetic fields $A_\mu(x)$ (for the simple example of such a procedure see, e.g., \cite{KazKor2024}). Keeping in mind that $p_c^\mu i\partial_\mu=im d/d\tau$ on particle's world line, it is easy to verify that the action functional,
\begin{equation}\label{action_IR}
    S[A_\mu(x),d_\mu(\tau)]:=-\frac14 \int d^4x F_{\mu\nu}(x)F^{\mu\nu}(x) +\int d\tau \big[-\frac{1}{8\pi r_0} \dot{d}^\mu_\perp \eta_{\mu\nu} \dot{d}^\nu_\perp +\dot{x}^\mu F_{\mu\nu}(x(\tau)) d^\nu \big],
\end{equation}
where $r_0:=\al/m$ is the classical electron radius, $d_\perp^\mu:=d^\mu-(\dot{x}d)\dot{x}^\mu$, $p_c^\mu=m\dot{x}^\mu$, and $\tau$ is the natural parameter on the world line, leads to such a system of equations. The action \eqref{action_IR} describes the interaction of the electric dipole moment $d_\mu$ with the electromagnetic field (see, e.g., \cite{rrmm}). This shows once again that, in the coherent processes, the wave packet of a single electron carries additional degrees of freedom -- the plasmons. In the infrared limit, these plasmons are reduced to the vector of the electric dipole moment. Notice that the classical description of the multipole moments of electron wave packets was investigated in \cite{KarlZhev19,PupKarl22}. The phenomenological approach introducing the polarizability of an electron has recently been considered in \cite{Jakobsen2024}.

\section{Conclusion}

Let us sum up the results. By using the $in$-$in$ formalism in the one-loop approximation, we have derived the explicit expression for the photon polarization operator in the presence of a single electron. The photon lines in this polarization operator are out of the mass-shell, while the electron state is described by the density matrix of a general form. The expression for the permittivity tensor of the electron wave packet easily follows from the polarization operator. On the photon mass-shell, the corresponding permittivity tensor coincides with the one obtained in \cite{KazSol2022} where it was established that the dielectric permittivity of a single electron wave packet is the same as the dielectric permittivity of a gas of free electrons. The results of the present paper reveal that this analogy keeps on out of the mass-shell.

We have investigated several approximations for the polarization tensor obtained. First, we have considered the polarization tensor in the case when the wavelength of an external electromagnetic field is much smaller than the typical scale of variation of the electronic wave packet in the coordinate space in the small quantum recoil limit. Then, for the electron wave packet sufficiently narrow in the momentum space, the polarization operator takes the form \eqref{polar_op_app}. Such a polarization operator coincides with the photon polarization operator in a gas of noninteracting electrons \cite{Lindhard1954,MelrWeis2003,Silin1960,Tsytovich1961,Melrose2008,VladTysh2011,Melrose2020}.

We have found the set of points in the momentum space of an off-shell photon where the polarization operator tends to infinity. As is known, such singularities manifest the presence of quasiparticles in the theory. In the case at issue, it is natural to call these quasiparticles plasmons despite the fact that, generally, plasmons describe collective excitations in a plasma \cite{TonLang1929,Lindhard1954,Pines1956,Pitarke2007,Bonitzb2015}. We have derived the dispersion law of plasmons \eqref{plasm_disp_law} propagating in a single electron wave packet. It is clear that the standard $in$-$in$ perturbation theory breaks down near these values of the off-shell photon momenta and the resummation of the perturbation series is needed. This resummation is achieved by going to the exact photon propagator instead of the free one. The exact photon propagator is standardly expressed in terms of the complete set of solutions of the effective Maxwell equations. We have obtained the solutions to the effective Maxwell equations and have found the eight independent plasmon-polariton modes. These eight plasmon-polaritons are supported by a single electron wave packet.

Second, we have derived the infrared asymptotics of the polarization tensor where the wavelength of the external electromagnetic field is much larger than the size of the electron wave packet. It turns out in this limit that the polarization tensor does not depend on the form of the wave packet and is the same as for a point dynamical electric dipole moving along the world line of the center of the electron wave packet. To put it another way, the plasmons discussed above are reduced to a dynamical electric dipole moment attached to the electron. This proves once more that a single electron carries additional effective degrees of freedom. We have found the action functional \eqref{action_IR} reproducing the effective Maxwell equations in the infrared limit.

The plasmons and the plasmon-polaritons supported by the electron wave packet become relevant for the coherent processes where the initial state of the electron coincides with the final one in the interaction representation. Such processes were studied in \cite{pra103,KazSol2022,KazSol2023,KRS2023,radet,BednNaum2021,KazinskiFr24} (see also \cite{MarcuseII,PanGov21,Talebi16,GovPan18,PanGov19,RoquesCarmes2023,Karnieli2024}). The simple example of such a coherent process is stimulated radiation from a single electron in an external electromagnetic field. The diagrams describing the contributions to the inclusive probability to record a photon are presented in Fig. \ref{Fig_Diagram}. The lines on this diagrams are the free vacuum propagators and the external lines of the customary $in$-$out$ perturbation theory. As is seen from the depicted diagrams, the initial and final states of the electron are the same and so the process is coherent. The diagrams shown in Fig. \ref{Fig_Diagram} describe the interference pattern of the incident photon beam with its scattered part recorded by the detector $D$ \cite{KazSol2022,KazSol2023,KRS2023,radet,KazinskiFr24}. This process resembles the construction of a classical hologram \cite{Gabor1949,SpBookMicrosc2019} where, in our case, the object is the electron wave packet. Just as for the classical hologram, the inclusive probability obtained in such experiments can be used to reconstruct the scattering amplitude. In the process depicted in Fig. \ref{Fig_Diagram}, there appear the diagrams describing the material contribution to the photon polarization operator we have calculated. They are the diagrams for the on-shell and off-shell coherent Compton process. The vacuum contribution is absent since in all these diagrams at least one of the photon lines is on the mass-shell. The contribution of these diagrams is large near the plasmonic resonances and the perturbation series has to be resummed as it has been described above. The presence of plasmonic resonances lead to enhancement of the amplitude of coherent scattering near them. The interaction of the single electron wave packet with the external electromagnetic field can be studied experimentally, for example, by confining the electron in a Penning trap \cite{Brown1986} or by many other means (see, e.g., \cite{RoquesCarmes2023,Karnieli2024}).

There are other coherent processes where the electron wave packet behaves as a charged fluid. We postpone a detailed investigation of the influence of plasmonic resonances on these processes for a future research.


\appendix
\section{Green's functions of the free electromagnetic field}\label{App_Green_Func_Bose}

In this appendix, we present the definitions of various Green's functions of the free electromagnetic field. In the $in$-$in$ perturbation theory, the following Green's functions arise
\begin{equation}\label{Green_in_in_A}
\begin{gathered}
    \lan A^\mu(x)\ran:=\lan\overline{in}|\hat{A}^\mu(x)|\overline{in}\ran,\\
    D^{\mu\nu}(x,y):=-i\lan\overline{in}|T\{\de\hat{A}^\mu(x) \de\hat{A}^\nu(y)\}|\overline{in}\ran=D^{\nu\mu}(y,x),\qquad D^{\mu\nu}_*(x,y):=i\lan\overline{in}|\tilde{T}\{\de\hat{A}^\mu(x)\de\hat{A}^\nu(y)\}|\overline{in}\ran,\\
    D^{\mu\nu}_{(+)}(x,y):=-i\lan\overline{in}|\de\hat{A}^\mu(x)\de\hat{A}^\nu(y)|\overline{in}\ran,\qquad D^{\mu\nu}_{(-)}(x,y):=i\lan\overline{in}|\de\hat{A}^\nu(y)\de\hat{A}^\mu(x)|\overline{in}\ran,\\
    D^{\mu\nu}_{(-)}(x,y)=\big[D^{\mu\nu}_{(+)}(x,y)\big]^* =-D^{\nu\mu}_{(+)}(y,x),
\end{gathered}
\end{equation}
where $\de\hat{A}^\mu(x):=\hat{A}^\mu(x)-\lan A^\mu(x)\ran$. Let us introduce the other Green's functions
\begin{equation}
\begin{gathered}
    D^{\mu\nu}_-(x,y):=-i\theta(x^0-y^0)[\hat{A}^\mu(x),\hat{A}^\nu(y)]=\theta(x^0-y^0)(D^{\mu\nu}_{(+)}(x,y)+D^{\mu\nu}_{(-)}(x,y)),\\ D^{\mu\nu}_+(x,y):=D^{\nu\mu}_-(y,x),\\
    \bar{D}^{\mu\nu}(x,y)=\frac12[D^{\mu\nu}_+(x,y) +D^{\mu\nu}_- (x,y)],\qquad D^{\mu\nu}_{(1)}(x,y)=i[D^{\mu\nu}_{(+)}(x,y) - D^{\mu\nu}_{(-)}(x,y)]=D_{(1)}^{\nu\mu}(y,x),
\end{gathered}
\end{equation}
where $D_-$, $D_+$, and $\bar{D}$ are the retarded, advanced, and symmetric Green's functions, respectively. They do not depend on the state $|\overline{in}\ran$. Using these definitions, one can verify that
\begin{equation}
    D=D_--D_{(-)}=D_++D_{(+)}=\bar{D}-\frac{i}{2}D_{(1)},\qquad D_*=D_--D_{(+)}=D_++D_{(-)}=\bar{D}+\frac{i}{2}D_{(1)}=D^*.
\end{equation}

\section{Other derivation of the expression for the polarization operator}\label{App_Other_Derivat}

Let us give in this appendix the other derivation of the expression \eqref{Pi_0}, \eqref{Pi_psi} for the polarization operator. This derivation allows one, in particular, to provide a more transparent physical interpretation to the matrix elements of polarization operator \eqref{polar_op_keld}. The Heisenberg equations for the quantum fields look as
\begin{equation}
    \Box \hat{A}_\mu(x)=e\hat{\bar{\psi}}(x)\ga_\mu\hat{\psi}(x),\qquad S_-^{-1}\hat{\psi}(x)=e\ga^\mu\hat{A}_\mu(x)\hat{\psi}(x).
\end{equation}
The graded Weyl-ordered ($C$-symmetric) expression for the operator of the current density is implied on the right-hand side of the quantum Maxwell equations. For conciseness, we do not write out this symmetrization explicitly. Using the standard asymptotic conditions, the quantum Dirac equation can be transformed into the integral equation
\begin{equation}
    \hat{\psi}(x)=\hat{\psi}_0(x)+e\int d^4y S_-(x,y)\ga^\mu\hat{A}_\mu(y)\hat{\psi}(y).
\end{equation}
In this appendix, as distinct from the other sections of the paper, $\hat{\psi}_0(x)$ is the Dirac field operator in the interaction representation and $\hat{\psi}(x)$ is the Dirac field operator in the Heisenberg representation. In the leading nontrivial order of the perturbation theory,
\begin{equation}
    \hat{\psi}(x)\approx\hat{\psi}_0(x)+e\int d^4y S_-(x,y)\ga^\mu\hat{A}_\mu(y)\hat{\psi}_0(y).
\end{equation}
Substituting this approximate expression into the the quantum Maxwell equation, we arrive at
\begin{equation}\label{Max_eqn}
    \Box \hat{A}_\mu(x) -e^2\int d^4y\big[\hat{\bar{\psi}}_0(x)\ga_\mu S_-(x,y)\ga^\nu\hat{A}_\nu(y)\hat{\psi}_0(y) +\hat{\bar{\psi}}_0(y)\ga^\nu\hat{A}_\nu(y)S_+(y,x)\ga_\mu\hat{\psi}_0(x) \big]=e\hat{\bar{\psi}}_0(x)\ga_\mu\hat{\psi}_0(x),
\end{equation}
where the graded Weyl ordering of the products of operators $\hat{\bar{\psi}}_0(x)$ and $\hat{\psi}_0(y)$ is understood.

Now it is clear that
\begin{equation}
    \Pi_{qc}^{\mu\nu}(x,y)=-e^2\lan\overline{in}|\hat{\bar{\psi}}_0(x)\ga^\mu S_-(x,y)\ga^\nu\hat{\psi}_0(y) +\hat{\bar{\psi}}_0(y)\ga^\nu S_+(y,x)\ga^\mu\hat{\psi}_0(x) |\overline{in}\ran,
\end{equation}
where, of course, the ultraviolet divergencies should be removed by the standard means in the vacuum contribution. To put it differently, in the given order of perturbation theory, the polarization operator $\Pi_{qc}$ is the operator acting on the field $\hat{A}_\mu$ in the quantum Maxwell equations, this operator being averaged over the initial state of the electron. In order to obtain the polarization operator $\Pi_{qq}$, one needs to evaluate
\begin{equation}
\begin{split}
    \lan\de\hat{A}_\mu(x)\de\hat{A}_\nu(y)\ran:=& \,\frac12\lan\overline{in}|\de\hat{A}_\mu(x)\de\hat{A}_\nu(y)+\de\hat{A}_\nu(y)\de\hat{A}_\mu(x)|\overline{in}\ran=\\ =&\,\frac12\lan\overline{in}|\hat{A}_\mu(x)\hat{A}_\nu(y)+\hat{A}_\nu(y)\hat{A}_\mu(x)|\overline{in}\ran-\bar{A}_\mu(x)\bar{A}_\nu(y),
\end{split}
\end{equation}
where
\begin{equation}
    \de\hat{A}_\mu(x):=\hat{A}_\mu(x)-\bar{A}_\mu(x),\qquad \bar{A}_\mu(x):=\lan\overline{in}|\hat{A}_\mu(x)|\overline{in}\ran,
\end{equation}
and the operator of the electromagnetic potential is found from \eqref{Max_eqn} in the leading order of perturbation theory
\begin{equation}
    \hat{A}_\mu(x)\approx \hat{A}^0_\mu(x)+e\int d^4y D_-(x,y)\hat{\bar{\psi}}_0(y)\ga_\mu\hat{\psi}_0(y).
\end{equation}
Here $D_-(x,y)$ is the retarded Green's function of the wave operator. Then it is not difficult to deduce that up to the terms of order $e^2$,
\begin{equation}
    \lan\de\hat{A}^\mu(x)\de\hat{A}^\nu(y)\ran=\frac12 \overset{(0)}{D}{}^{\mu\nu}_{(1)}(x,y) -i\int d^4x_1 d^4y_1 D_-(x,x_1) D_-(y,y_1) \Pi^{\mu\nu}_{qq}(x_1,y_1),
\end{equation}
where (see Appendix \ref{App_Green_Func_Bose})
\begin{equation}
    \overset{(0)}{D}{}^{\mu\nu}_{(1)}(x,y):=\lan0|\hat{A}_0^\mu(x)\hat{A}_0^\nu(y) +\hat{A}_0^\nu(y)\hat{A}_0^\mu(x)|0\ran.
\end{equation}
We see that the polarization operator $\Pi_{qq}$ specifies the quantum corrections to fluctuations of the electromagnetic field.

\section{Tensors $G^{\mu\nu}(\spp,\spp')$ and $Z^{i\mu\nu}(\spp,\spp')$}\label{App_G_Z}

In this appendix, we give the explicit complete expressions for the tensors $G^{\mu\nu}(\spp,\spp')$ and $Z^{i\mu\nu}(\spp,\spp')$ appearing in the decomposition of the polarization operator \eqref{Pi_psi_ex1}. After a little algebra, we obtain
\begin{equation}
    G^{\mu\nu}(\spp,\spp')=\frac{1}{m\sqrt{(p^0_c+m)^2-q_0^2/4}}\Big[\frac{g^{\mu\nu}}{(p^c-k_+^c)^2-m^2} +(\mu\leftrightarrow\nu,k_c^\mu\rightarrow-k_c^\mu)\Big],
\end{equation}
where
\begin{equation}
\begin{split}
    g^{\mu\nu}:=&\,\frac12 \eta^{\mu\nu}\big[(k_cp_c)(p^0_c+m) +\frac14(k_c^0q^2-(k_cq)q^0) \big] +(p_c^0+m)p_c^\mu p_c^\nu-\\
    &-p^\mu_c \big[ (p_c^0+m)\big(k_c^\nu+\frac{q^\nu}{2}\big) +\frac{\de_0^\nu}{2} \big((k_cq)-\frac{q^2}{2}\big) -\frac{q^\nu}{2}\big(k_c^0-\frac{q^0}{2}\big) \big] -\frac14\de_0^\mu\big[ k_c^\nu q^2 +2q^\nu\big((k_cp_c)+\frac{q^2}{4}\big)\big]+\\
    &+\frac14q^0k_c^\mu q^\nu+(\mu\leftrightarrow\nu,q^\mu\rightarrow-q^\mu).
\end{split}
\end{equation}
As for the second tensor, we have
\begin{equation}
    \xi_iZ^{i\mu\nu}(\spp,\spp')=\frac{i}{4m\sqrt{(p^0_c+m)^2-q_0^2/4}}\Big[\frac{z^{\mu\nu}}{(p^c-k_+^c)^2-m^2} +(\mu\leftrightarrow\nu,k_c^\mu\rightarrow-k_c^\mu)\Big],
\end{equation}
where
\begin{equation}
\begin{split}
    z^{\mu\nu}:=&\,4\e^{\mu\nu\rho\la}k^c_\rho\xi_\la m(p^c_0+m) -4\e^{\mu\nu\rho\s}(p_c\xi)k^c_\rho (p^c_\s+m\de^0_\s) +2\eta^{\mu\nu}\e^{\rho\s\la\vk}k^c_\rho(p^c_\s+m\de^0_\s)q_\la\xi_\vk+\\
    &+2(p_c^{(\mu} -k_c^{(\mu})\e^{\nu)\rho\s\la}(p^c_\rho+m\de^0_\rho) q_\s\xi_\la +\e^{\mu\nu\rho\la} (p^c_\rho-k^c_\rho) (q^2\xi_\la-(q\xi) q_\la) +mq^{[\mu}\e^{\nu]0\rho\la}q_\rho\xi_\la,
\end{split}
\end{equation}
and $\xi^\la=(0,\bs\xi)$. Notice that $G^{\mu\nu}(\spp,\spp')$ is symmetric under the replacement $\mu\leftrightarrow\nu$ and $q^\mu\rightarrow-q^\mu$, whereas the tensor $Z^{i\mu\nu}(\spp,\spp')$ changes its sign under this substitution.



\end{document}